\documentclass[amsmath,amssymb,eqsecnum,notitlepage,nofootinbib,a4paper]{revtex4-1}
\usepackage{graphicx}
\usepackage{fontenc}
\usepackage{subfigure}
\usepackage{setspace}
\usepackage[breaklinks, colorlinks, citecolor=magenta, linkcolor=blue!90!black]{hyperref}
\usepackage{amsmath}
\usepackage{amssymb}
\usepackage{bm}% bold math
\usepackage{verbatim}
\usepackage{xcolor}

\graphicspath{{./}}

\def\half{\frac{1}{2}}
\def\third{\frac{1}{3}}
\def\quarter{\frac{1}{4}}
\def\0{\varnothing} % component zeroth order in spatial derivatives, time derivatives and momenta
\def\T{\mathsf{T}} % traceless component
\def\BI{\gamma_{\mathrm{BI}}} % Barbero-Immirzi parameter

\def\R{\mathcal{R}}
\def\K{\mathcal{K}}

\def\H{\mathcal{H}}
\def\bp{\mathcal{P}}

\def\xty{\left({x}\leftrightarrow{y}\right)}
\def\gud{\mathrm{gd}}
\def\sech{\mathrm{sech}}

\newcommand{\four}[1]{{}^{(4)}\!{#1}}
\newcommand{\partdif}[2]{\frac{\partial {#1}}{\partial {#2}}}
\newcommand{\funcdif}[2]{\frac{\delta {#1}}{\delta {#2}}}

\newcommand{\figref}[1]{Fig.~\ref{#1}}

\newcommand{\secref}[1]{section~\ref{#1}}

\newcommand{\appref}[1]{appendix~\ref{#1}}
\newcommand{\sgn}[1]{\operatorname{sgn} (#1)}

\begin{document}

\begin{flushleft}
KCL-PH-TH/2018-79
\end{flushleft}

\title{The general scalar-tensor Hamiltonian with deformed covariance}

\author{Rhiannon Cuttell} \email{rhiannon.cuttell@kcl.ac.uk}
\author{Mairi Sakellariadou} \email{mairi.sakellariadou@kcl.ac.uk}
\affiliation{
    Department of Physics,
	King's College London,
	University of London,
	Strand,
	London,
	WC2R 2LS,
	U.K.
}
\date{\today}

\begin{abstract}
We derive the scalar-tensor Hamiltonian constraint to all orders of momenta when the canonical constraint algebra is deformed by a phase space function as predicted by some studies into loop quantum cosmology.
We find that the momenta and spatial derivatives from gravity and matter fields must combine in a very specific form to maintain a deformed general covariance.  This suggests that the deformation should be equally caused by matter field derivatives as it is by gravitational curvature.
We demonstrate the cosmological consequences of a variety of deformation functions.  Some cause a big bounce at high energy densities and others cause a sudden singularity.
\end{abstract}

\maketitle

%%%%%%%%%%%%%%%%%%%%%%%%%%%%%%%%%%%%%%%%%%%%%%%%%%%%%%%%%%%%%%%%%%%%%%%%%%

\tableofcontents % Prints the main table of contents

%%%%%%%%%%%%%%%%%%%%%%%%%%%%%%%%%%%%%%%%%%%%%%%%%%%%%%%%%%

\section{Introduction}
\label{sec:intro}

% Something about the importance of general covariance (and what it is)

One of the enduring insights of Einstein's general relativity is that physics should ideally be generally covariant.  That is, there should be no preferred observer or reference frame.
However, this means that attempts to reconcile general relativity and quantum mechanics inevitably must deal with the problem of time \cite{Rovelli:1989jn, Isham:1992ms}.  In quantum mechanics time is a fixed external parameter, in general relativity it is internal to the system and is not uniquely defined.

% When using canonical methods, space and time are distinguished, and each dimension has a conserved quantity or `constraint'

% general covariance is assured by the structure of the poisson algebra of the constraints

The solution in canonical gravity for reconciling the two is to split space-time at the formal level, but include symmetry requirements so that the full general covariance is kept implicitly \cite{Arnowitt:1962hi, Gourgoulhon:2007ue, bojowald2010canonical}.  One is left with a description of a spatial slice evolving through time rather than one of a static and eternal bulk.
These methods are often required for numerically simulating general relativity due to the necessity of specifying a time coordinate when setting up an evolution simulation.

\begin{sloppypar}
This introduces on each spatial manifold a conserved quantity or `constraint' given by ${\phi_I\to0}$ for each dimension of time and space, analogous to the conservation of energy and momentum. 
These constraints form a Poisson algebra which contains important information about the geometric nature of space-time, and is of the form ${\left\{\phi_I,\phi_J\right\}=f_{IJ}^K\phi_K}$ \cite{dirac1964, bojowald2010canonical}.
This is a Lie algebroid which describes the relationships between the constraints and generates transformations between different choices of coordinates \cite{hojman_geometrodynamics_1976, Bojowald:2016hgh}.
\end{sloppypar}

%%%%%%%%%%%%%%%%%%%%%%%%%%%%%%%%%%

For models of loop quantum cosmology to be self-consistent and anomaly-free while including some of the interesting effects from the discrete geometry, it seems that the algebra of constraints must be deformed.  Specifically, some of the structure functions become more dependent on the phase space variables through a deformation function ${f_{IJ}^K(q)\to{}\beta(q,p)f_{IJ}^K(q)}$ \cite{Bojowald:2008gz, Bojowald:2008bt, Perez:2010pm, Mielczarek:2011ph, Cailleteau2012a, Mielczarek:2012pf, Cailleteau2013}.  Deforming rather than breaking the algebra in principle maintains general covariance but the transformations between different choices of reference frame become highly non-linear \cite{tibrewala_inhomogeneities_2013}.  It becomes less clear to what extent one can still interpret space-time geometrically, at least in terms of classical notions of geometry.

However, there is ambiguity in the correct choice of variables used for loop quantum gravity.  The results cited in the previous paragraph are for real variables for which there has been significant difficulty including matter and local degrees of freedom \cite{Bojowald:2016itl}.  The main alternative, self-dual variables, have had some positive results for including those degrees of freedom without deforming the constraint algebra \cite{BenAchour:2016brs}, but might not have the desired quality of resolving curvature singularities \cite{Brahma:2016tsq}.

%%%%%%%%%%%%%%%%%%%%%%%%%%%%%%%%%%%%%%

Deformed general relativity is a semi-classical model built directly from the idea that the constraint algebra is deformed \cite{bojowald_deformed_2012}. It is constructed by taking the deformed constraint algebra, and finding a corresponding model which includes local degrees of freedom \emph{a priori}.
This can be done because, if one starts from an algebra and makes some reasonable assumptions, one can deduce the general form of all the constraints \cite{kuchar_geometrodynamics_1974, hojman_geometrodynamics_1976}.  This should provide a more intuitive understanding of how the deformation affects dynamics and may provide a guide for how to include the problematic degrees of freedom when working with real variables in loop quantum cosmology.

The constraint algebra is important because, as said previously, it closely relates to the structure of space-time \cite{teitelboim_how_1973}.  
Quantum geometry will behave differently to classical geometry, and deformed general relativity attempts to capture some of the effects in a semi-classical model which is more amenable to phenomenological investigations.

Phenomenological models which are comparable to deformed general relativity, such as deformed special relativity \cite{AmelinoCamelia:2000mn} and rainbow gravity \cite{Magueijo:2002xx}, struggle to go beyond describing individual particles coupled to an energy-dependent metric.  They can suffer from a breakdown of causality \cite{hossenfelder_box-problem_2009}, or find it difficult to describe multi-particle states \cite{hossenfelder_multi-particle_2007}.  Deformed general relativity does not suffer from these problems by construction.

In this paper, we seek to find the most general Hamiltonian constraint for a scalar-tensor model which satisfies the deformed constraint algebra.  We find the restrictions on the form of the constraint in \secref{sec:dist-eqn}, we use them to derive the constraint in \secref{sec:allst}, and we investigate some of the cosmological implications in \secref{sec:cosmo}. 
These calculations generalise those presented in ref.~\cite{cuttell2018} where the minimally deformed scalar-tensor constraint was derived from the constraint algebra.
For a more in-depth review, please see ref.~\cite{cuttellthesis}.

%%%%%%%%%%%%%%%%%%%%%%%%%%%%%%%%%%%%%%

\subsection{Space-time decomposition}
\label{sec:methodology_decomposition}

Quantum mechanics naturally works in the canonical or Hamiltonian framework.
The canonical framework takes variables defined at a certain time and evolves them through time.  That evolution defines a canonical momentum for each variable.
To make general relativity more amenable to quantum mechanics, one must likewise make a distinction between the time dimension and the spatial dimensions.
So we foliate the bulk space-time manifold $\mathcal{M}$ into a stack of labelled spatial hypersurfaces, $\Sigma_t$. We assume no closed causal curves, so topologically $\mathcal{M}=\Sigma\times\mathbb{R}$ \cite{bojowald2010canonical, Arnowitt:1962hi, Gourgoulhon:2007ue}.

The time-evolution is described by the time vector $t^a$.
Each spatial slice has a metric $q_{ab}$, and a future-pointing normal vector $n^a$, so we can project the time vector into its tangential and normal components.  This produces the lapse function ${N=-n_{a}t^{a}}$ and the spatial shift vector ${N^{a}=q^{a}_{b}t^{b}}$.  These act as Lagrange multipliers in the classical action, and in the canonical formalism they produce constraints from the total Hamiltonian,
\begin{equation}
    C : = \funcdif{ H }{ N },
\quad
    D_a : = \funcdif{ H }{ N^a },
\end{equation}
which are respectively known as the Hamiltonian constraint, and the diffeomorphism constraint. 

For a canonical system, the constraints $\phi_I$ have a Poisson bracket structure,
\begin{equation}
    \{ \phi_I, \phi_J \} = f_{IJ}^K \phi_K + \alpha_{IJ},
\quad
    \alpha_{IJ} \notin \{ \phi_K \},
    \label{eq:con-alg_general}
\end{equation}
and if there are anomalous terms, $\alpha_{IJ}\neq0$, then some of $\phi_I$ are what are called `second-class' constraints, in which case some of the Lagrange multipliers of the system are uniquely determined.  If $\alpha_{IJ}=0$ then all of $\phi_I$ are `first-class', in which case the constraints not only restrict the values of the dynamical fields, but also generate gauge transformations \cite{dirac1964, bojowald2010canonical}.  This is because, in general the equations of motion depend on the Lagrange multipliers.  For an undetermined multiplier to influence the mathematics but not the physical observables, a change of its value must correspond to a gauge transformation generated by the relevant first-class constraint.

For classical general relativity in the canonical formalism, the Poisson bracket structure \eqref{eq:con-alg_general} of the constraints ${\phi_I\in\{C,D_a\}}$ forms a Lie algebroid%
\footnote{`Algebroid' refers to the fact that some of the structure coefficients $f^K_{IJ}$ are phase space functions}\cite{Bojowald:2016hgh},
\begin{subequations}
\begin{align}
    \big\{ D_a [N^a], D_b [M^b] \big\} & = D_a \big[ \mathcal{L}_M N^a \big],
        \label{eq:con-alg_DD}
        \\
    \big\{ C [N], D [M^a] \big\} & = C \big[ \mathcal{L}_M N \big],
        \label{eq:con-alg_CD}
        \\
    \big\{ C [N], C [M] \big\} & = D_a \big[ \, q^{ab} \left( N \partial_b M - \partial_b N M \right) \big].
        \label{eq:con-alg_CC}
\end{align}
    \label{eq:con-alg}%
\end{subequations}
Since there are no anomalous terms, $N$ and $N^a$ are gauge functions which do not affect the observables, and therefore the spatial slicing does not affect the dynamics.  
The theory is background independent and the constraints generate the transformations%
\footnote{The square brackets indicates the constraint is `smeared' over the spatial surface using the function in the brackets, e.g. ${C[N]=\int\mathrm{d}^3xN(x)C(x)}$.  The Lie derivative of $F$ with respect to the vector $N^a$ is given by ${\mathcal{L}_{N}F}$.},
\begin{equation}
    \{ F, C [ N ] \} = N \mathcal{L}_n F,
\quad
    \{ F, D_a [ N^a ] \} = \mathcal{L}_N F.
    \label{eq:methodology_gauge_transformations}
\end{equation}
As interpreted in ref.~\cite{teitelboim_how_1973}, \eqref{eq:con-alg_DD} shows that $D_a$ is the generator of spatial morphisms, \eqref{eq:con-alg_CD} shows that $C$ is a scalar density of weight one, and \eqref{eq:con-alg_CC} specifies the form of $C$ such that it ensures the embeddability of the spatial slices in space-time geometry.

%%%%%%%%%%%%%%%%%%%%%%%%%%%%%%%%%%%%%%%%%%%%%%%%%%%%%%%%%%%%%%%%%

\subsection{Choice of variables}
\label{sec:methodology_variables}

Classical canonical gravity can be formulated equivalently using different variables.  There is geometrodynamics, which uses the spatial metric and its canonical momentum $(q_{ab}, p^{cd})$, the latter of which is directly related to extrinsic curvature ${K_{ab}=\half\mathcal{L}_{n}q_{ab}}$,
\begin{equation}
    p^{ab} = \frac{ \omega }{ 2 } \sqrt{ q } \left( K^{ab} - K q^{ab} \right),
\end{equation}
where $q:=\det{q_{ab}}$, and $\omega=1/8\pi{G}$ is the gravitational coupling.
An alternative is connection dynamics, which uses the Ashtekar-Barbero connection and densitised triads $(A^I_a, E^b_J)$, where capital letters signify internal indices rather than coordinate indices \cite{Ashtekar:1986yd, Barbero:1994ap}. This can be related to geometrodynamics by using the equations \cite{bojowald2010canonical},
\begin{equation}
    q \, \delta_{IJ} = q_{ab} E^a_I E^b_J ,
\quad
    A_a^I = \Gamma_a^I + \BI K_a^I ,
\quad
    \Gamma_a^I = \frac{ 1 }{ 2 \sqrt{ q } } q_{bc} \epsilon^{IJK} E^b_J \nabla_a \left( \frac{ E^c_{K} }{ \sqrt{ q } } \right) ,
\quad
    K_a^I = \frac{ 1 }{ \sqrt{ q } } \delta^{IJ} K_{ab} E^{b}_J ,
\end{equation}
where $\BI$ is the Barbero-Immirzi parameter and $\epsilon^{IJK}$ is the covariant Levi-Civita tensor.  The exact value of $\BI$ should not affect the dynamics \cite{Immirzi:1996dr}.

The main alternative is loop dynamics, which uses holonomies of the connection and gravitational flux $(h_{\ell}[A],F^I_{\ell}[E])$.  Classically, $h_{\ell}[A]$ is given by the path-ordered exponential of the connection integrated along a curve ${\ell}$ and $F^I_{\ell}[E]$ is the flux of the densitised triad through a surface that the curve ${\ell}$ intersects. If ${\ell}$ is taken to be infinitesimal, one can easily relate loop dynamics and connection dynamics \cite[p.~21]{Rovelli:2014ssa}.

When each set of variables is quantised, they are no longer equivalent, for example the value of $\BI$ does now affect the dynamics \cite{Ashtekar:1997yu, Brahma:2016tsq}. For complex $\BI$, care has to be taken to make sure the classical limit is real general relativity, rather than complex general relativity.  Significantly, quantising loop variables (loop quantum gravity) discretises geometry, and so $\ell$ cannot be taken to be infinitesimal \cite[p.~105]{Rovelli:2014ssa}.

In this work, we choose to use metric variables to build a semi-classical model of gravity. This is because the comparison to other modified gravity models should be clearer, and there is no ambiguity like that which arises from $\BI$.

For simplicity, we are not including degrees of freedom beyond a simple scalar-tensor model.  Since actions which contain Riemann tensor squared contractions such as
${\four{R}^{ab}\four{R}_{ab}}$
and 
${\four{R}^{abcd}\four{R}_{abcd}}$
introduce additional tensor degrees of freedom \cite{Deruelle2010}, we automatically do not consider related terms here.
This means we only need to expand the constraint using variables which are tensor contractions containing up to two orders of spatial derivatives or up to two orders in momenta.  It also means we do not need to consider spatial derivatives of momenta in the constraint.
Therefore, for a metric tensor field ${\left(q_{ab},p^{cd}\right)}$ and a scalar field ${\left(\psi,\pi\right)}$, we expand the constraint into the following variables,
\begin{equation}
\begin{gathered}
    q = \det{q_{ab}},
        \quad
    p = q_{ab} p^{ab},
        \quad
    \bp = Q_{abcd} p_\T^{ab} p_\T^{cd},
        \quad
    R,
        \\
    \psi,
        \quad
    \pi,
        \quad
    \Delta := q^{ab} \nabla_a \nabla_b \psi 
    = \partial^2 \psi - q^{ab} \Gamma^c_{ab} \partial_c \psi,
        \quad
    \gamma := q^{ab} \nabla_a \psi \nabla_b \psi = \partial^a \psi \partial_a \psi,
\end{gathered}
        \label{eq:allst_variables}
\end{equation}
where 
${p_\T^{ab}:=p^{ab}-\third{}pq^{ab}}$
is the traceless part of the metric momentum.
Therefore, we use the constraint given by ${C=C(q,p,\bp,R,\psi,\pi,\Delta,\gamma)}$.

%%%%%%%%%%%%%%%%%%%%%%%%%%%%%%%%%%%%%%%%%%%%%%%%%%%%%%%%%%%%%%%%%%%%%%%

\subsection{Deformed constraint algebra}

Some models of loop quantum cosmology predict that the symmetries of general relativity should be deformed in a specific way in the semi-classical limit \cite{Bojowald:2008gz, Bojowald:2008bt, Perez:2010pm, Mielczarek:2011ph, Cailleteau2012a, Mielczarek:2012pf, Cailleteau2013}.  This appears from incorporating loop variables in a mini-superspace model, but specifying that all anomalies $\alpha_{IJ}$ in \eqref{eq:con-alg_general} vanish while allowing counter-terms to deform the classical form of the algebra.
This ensures that the constraints are first-class, retaining the gauge invariance of the theory and of the arbitrariness of the lapse and shift.  
If anomalous terms \emph{were} to appear in the constraint algebra, then the gauge invariance would be broken and the constraints could only be solved at all times for specific $N$ or $N^a$.  This means that there would a privileged frame of reference, and therefore no general covariance.

In the referenced studies, it is strongly indicated that the bracket of two Hamiltonian constraints \eqref{eq:con-alg_CC} is deformed by a phase space function $\beta$,
\begin{equation}
    \{ C [N] , C [M] \} = D_a [ \beta q^{ab} \left( N \partial_b M - \partial_b N M \right) ].
    \label{eq:con-alg_def}
\end{equation}
This has not been shown generally, but has been shown for several models independently.
There are no anomalies in the constraint algebra, so a form of general covariance is preserved.  However, it may be that the interpretation of a spatial manifold evolving with time being equivalent to a foliation of space-time (also known as `embeddability') is no longer valid.

These deformations only appear to be necessary for models when the Barbero-Immirzi parameter $\BI$ is real.  For self-dual models, when $\BI=\pm{i}$, this deformation does not appear necessary \cite{BenAchour:2016brs}.  However, self-dual variables are not desirable in other ways.  They do not seem to resolve curvature singularities as hoped, and obtaining the correct classical limit is non-trivial \cite{Brahma:2016tsq}.  Because of this, even though we use metric variables in this study, considering $\beta\neq1$ and ensuring the correct classical limit means there should be relevance to the models of loop quantum cosmology with real $\BI$.

%%%%%%%%%%%%%%%%%%%%%%%%%%%%%%%%%%%%%%%%%%%%%%%%%%

\subsection{Derivation of the distribution equation}
\label{eq:methodology_dist-eqn}

From the constraint algebra, we are able to find the specific form of the Hamiltonian constraint $C$ for a given deformation $\beta$.  The diffeomorphism constraint $D_a$ is not affected when the deformation is a scalar of weight zero, and so is completely determined as shown in \appref{sec:diff}.  With $D_a$ and $\beta$ as inputs, we can find $C$ by manipulating \eqref{eq:con-alg_def}.

Firstly, we must find the unsmeared form of the deformed algebra.  At this point we do not need to specify our canonical variables, and leave them merely as $\left(q_I,p_I\right)$,
\begin{subequations}
\begin{align}
    0 & = \Big\{ C [N] , C [M] \Big\} - D_a \Big[ \beta q^{ab} \left( N \partial_b M - \partial_b N M \right) \Big],
\\
    & = \int \mathrm{d}^3 z \left\{ 
        \sum_I \funcdif{ C [N] }{ q_I(z) } \funcdif{ C [M] }{ p_I (z) }
        - \left( \beta D^a N \partial_a M \right)_z
    \right\}
    - \left( N \leftrightarrow M \right).
    \label{eq:con-alg_def_expanded}
\end{align}
\end{subequations}
Take the functional derivatives with respect to $N(x)$ and $M(y)$, and note that we will only consider constraints without spatial derivatives on momenta,
\begin{equation}
    0 = \sum_I \funcdif{ C (x) }{ q_I (y) } \left. \partdif{ C }{ p_I } \right|_y
    - \left( \beta D^a \partial_a \right)_x \delta \left( x, y \right) 
    - \xty.
    \label{eq:dist-eqn_con}
\end{equation}
This is the key equation used as a basis for finding the constraint for deformed general relativity.

%%%%%%%%%%%%%%%%%%%%%%%%%%%%%%%%%%%%%%%%%%%%%%%%%%%%%%%%%%%%%%%%%%%%%%%

\subsection{Order of the deformed constraint}
\label{sec:methodology_order}

We can determine the relationship between the order of the deformation function and the order of the associated constraint by comparing orders of momenta .
As an example, take the distribution equation \eqref{eq:dist-eqn_con} with only a scalar field,
\begin{equation}
    0 = \funcdif{ C (x) }{ \psi (y) } \left. \partdif{ C }{ \pi } \right|_y
    - \left( \beta \pi \partial^a \psi \partial_a \right)_x \delta \left( x, y \right)
    - \xty,
    \label{eq:dist-eqn_con_scalar}
\end{equation}
where we have used the diffeomorphism constraint \eqref{eq:diff_scalar}.
we take a simplified model with two spatial derivatives represented by $\Delta$, only taking even orders of derivatives because of assuming spatial parity.  We take the distribution equation \eqref{eq:dist-eqn_con_scalar} and put it into schematic form,
\begin{equation}
    0 = \partdif{ C }{ \Delta } \partdif{ C }{ \pi } 
    - \beta \, \pi.
    \label{eq:schematic_constraint_all}
\end{equation}
so that we can consider orders of $\pi$ in a way analogous to dimensional analysis.
This equation must be satisfied independently at each order of momenta, so we isolate the coefficient of $\pi^n$,
\begin{equation}
    0 = \sum_{m=1}^{n_C} m \partdif{ C^{(n-m+1)} }{ \Delta } C^{(m)}
    - \beta^{(n-1)},
    \label{eq:schematic_constraint_coefficient}
\end{equation}
where we have expanded the constraint and deformation, 
\begin{equation}
    C=\sum_{m=0}^{n_C}C^{(m)}\pi^m,
\quad
    \beta=\sum_{m=0}^{n_\beta}\beta^{(m)}\pi^m.
\end{equation}
The highest order contribution to \eqref{eq:schematic_constraint_coefficient} comes when ${m=n_C}$ and ${n-m+1=n_C}$, in which case ${n=2n_C-1}$.  This is the highest order at which $\beta$ won't automatically be constrained to vanish, so we find its highest order of momenta to be ${n_\beta=2n_C-2}$.
However, this result does not take into account the fact that the combined order of momenta and spatial derivatives may be restricted.  If this is the case (as is found in \secref{sec:allst}), then the highest order contribution to the \eqref{eq:schematic_constraint_coefficient} will be when ${n-m+1=n_C-2}$, in which case we find the relation
\begin{equation}
    2 n_C - n_\beta = 4 .
    \label{eq:methodology_order_constraint}
\end{equation}
We see that a deformed second order constraint only requires considering a zeroth order deformation as in ref.~\cite{cuttell2018}, but a fourth order constraint requires considering a fourth order deformation.  We consider the constraint to general order in this paper.  Note that this relation suggests there are higher order deformations which allow for constraints given by finite order polynomials, unlike the generally deformed action in ref.~\cite{cuttellaction} which does not seem to permit a polynomial solution.

%%%%%%%%%%%%%%%%%%%%%%%%%%%%%%%%%%%%%%%%%%%%%%%%%%%%%%%%%

\subsection{Cosmology}
\label{sec:methodology_cosmo}

Since the main motivations for this study centre around cosmological implications of the deformed constraint algebra, we need to lay out how to find the cosmological dynamics of a model.
We restrict to an isotropic and homogeneous space, using the Friedmann-Lema\^{i}tre-Robertson-Walker metric (FLRW) given by
${q_{ab}=a^2\Sigma_{ab}}$,
and
${N^a=0}$,
where
${a:=\left(\det{q_{ab}}\right)^{1/6}}$,
and
$\Sigma_{ab}$ is the spatial metric of a static spatial slice with constant curvature $k$.
The normal derivative of the spatial metric is given by,
${v_{ab}=2a\dot{a}N^{-1}\Sigma_{ab}=:2a^2{\H}\Sigma_{ab}}$,
where $\H$ is the Hubble expansion rate, and the Ricci curvature scalar is given by
${R=6ka^{-2}}$.
When using canonical coordinates, the metric momentum is given by
${p^{ab}=\bar{p}\Sigma^{ab}}$,
where,
${\bar{p}=\left(\det{p^{ab}}\right)^{1/3}}$,
which changes the metric's commutation relation,
\begin{equation}
    \Big\{ q_{ab} (x), p^{cd} (y) \Big\} = \delta_{ab}^{cd} (x) \delta ( x, y )
\quad\to\quad
    \Big\{ a (x), \bar{p} (y) \Big\} = \frac{ \delta ( x, y ) }{ 6 a (x) } .
\end{equation}
The spatial derivatives of matter fields vanish, $\partial_a \psi_I = 0$.
One may couple a perfect fluid to the metric by including the energy density $\rho$ in the constraint \cite{Brown:1992kc},
${C\supset{a}^3\rho}$,
which must satisfy the continuity equation,
\begin{equation}
    \dot{\rho} + 3 \H \rho \left( 1 + w \right) = 0,
        \label{eq:methodology_cosmo_continuity}
\end{equation}
where $w$ is the perfect fluid's equation of state, the ratio of the pressure density to the energy density.

Ideally, we would like to extract cosmological observables such as the primordial scalar index to find phenemenological constraints.  However, to calculate the power spectra of primordial fluctuations would require adapting the cosmological perturbation theory formalism to ensure it is valid for deformed general covariance.  This would probably be highly non-trivial and therefore has been left for possible future study.

%%%%%%%%%%%%%%%%%%%%%%%%%%%%%%%%%%%%%%%%%%%%%%%%%%%%%%%%%%

\section{Solving the distribution equation}
\label{sec:dist-eqn}

From \eqref{eq:dist-eqn_con}, we have the general distribution equation for a Hamiltonian constraint, without derivatives of the momenta, which depends on a metric tensor and a scalar field,
\begin{equation}
    0 = 
    \funcdif{ C (x) }{ q_{ab} (y) } 
    \left. \partdif{ C }{ p^{ab} } \right|_y 
    + \funcdif{ C (x) }{ \psi (y) }
    \left. \partdif{ C }{ \pi } \right|_y
    - \left( \beta D^a \partial_a \right)_x \delta \left( x, y \right) - \left( x \leftrightarrow y \right).
        \label{eq:allst_dist-eqn}
\end{equation}
To solve this we will take the functional derivative with respect to a momentum variable, perform some algebraic manipulation, and then integrate with a test tensor to find several equations which the constraint must satisfy.  Since we have two fields, we must do this procedure twice.  The first route we consider will be where we take the derivative with respect to the metric momentum $p^{ab}$.  We will follow this by the route related to the scalar field momentum $\pi$.

\subsection{\texorpdfstring{$p^{ab}$}{Metric momentum} route}
\label{sec:dist-eqn-sol-metric}

Starting from the distribution equation \eqref{eq:allst_dist-eqn}, relabel indices, then take the functional derivative with respect to $p^{ab}(z)$,
\begin{equation}
\begin{split}
    0 & =
    \funcdif{ C (x) }{ q_{cd} (y) } \left. \partdif{^2 C }{ p^{ab} \partial p^{cd} } \right|_y \delta ( z, y )
    + \funcdif{ \partial C (x) }{ q_{cd} (y) \partial p^{ab} (x) } \left. \partdif{ C }{ p^{cd} } \right|_y \delta ( z, x )
    + \funcdif{ C (x) }{ \psi (y) } \left. \partdif{^2 C }{ p^{ab} \partial \pi } \right|_y \delta ( z, y )
\\ &
    + \funcdif{ \partial C (x) }{ \psi (y) \partial p^{ab} (x) } \left. \partdif{ C }{ \pi } \right|_y \delta ( z, x )
    - \partial_{c(x)} \delta ( x, y ) 
    \left(
        \partdif{ ( \beta D^c ) }{ p^{ab} } + \beta \partdif{ D^c }{ p^{ab}_{,d} } \partial_d
    \right)_x \delta ( z, x)
    - \xty.
\end{split}
\end{equation}
Move derivatives and discard surface terms so that it is
reorganised into the form,
\begin{equation}
    0 = 
    A_{ab} ( x, y ) \delta ( z, y )
    - A_{ab} ( y, x ) \delta ( z, x ),
    \label{eq:allst_dist-eqn_Aab}
\end{equation}
where,
\begin{equation}
\begin{split}
    A_{ab} ( x, y ) & =
    \funcdif{ C (x) }{ q_{cd} (y) } \left. \partdif{^2 C }{ p^{ab} \partial p^{cd} } \right|_y
    - \funcdif{ \partial C (y) }{ q_{cd} (x) \partial p^{ab} (y) } \left. \partdif{ C }{ p^{cd} } \right|_x
    + \funcdif{ C (x) }{ \psi (y) } \left. \partdif{^2 C }{ p^{ab} \partial \pi } \right|_y
\\ &
    - \funcdif{ \partial C (y) }{ \psi (x) \partial p^{ab} (y) } \left. \partdif{ C }{ \pi } \right|_x
    + \left( \partdif{ ( \beta D^c ) }{ p^{ab} } \partial_c \right)_y \delta ( y, x )
    - \partial_{d(y)} \left\{ \left( \beta \partdif{ D^c }{ p^{ab}_{,d} } \partial_c \right)_y \delta ( y, x ) \right\}.
\end{split}
    \label{eq:allst_Aab}
\end{equation}
If we take \eqref{eq:allst_dist-eqn_Aab} and integrate over $y$, we can find $A_{ab}(x,y)$ in terms of a function dependent on only a single independent variable,
\begin{equation}
    0 = A_{ab} ( x, z ) - A_{ab} (x) \delta ( z, x ),
    \quad
    \mathrm{where,} \;
    A_{ab} (x) = \int \mathrm{d}^3 y A_{ab} ( y, x ).
\end{equation}
We then multiply this by an arbitrary, symmetric test tensor $\theta^{ab}(z)$, integrate over $z$, and separate out different orders of derivatives of $\theta^{ab}$,
\begin{equation}
\begin{split}
    0 & =
    \theta^{ab} \! \left( \cdots \right)_{ab}
    + \partial_c \theta^{ab} \! \left\{
        \partdif{ C }{ q_{ef,c} } \partdif{^2 C }{ p^{ab} \partial p^{ef} }
        + 2 \partdif{^2 C }{ q_{ef,cd} } \partial_d \left( \! \partdif{^2 C }{ p^{ab} p^{ef} } \! \right)
        + \partdif{ C }{ p^{ef} } \partdif{^2 C }{ q_{ef,c} \partial p^{ab} }
        - 2 \partdif{ C }{ p^{ef} } \partial_d \left( \! \partdif{^2 C }{ q_{ef,cd} \partial p^{ab} } \! \right)
\right. \\ & \left.
        + \partdif{ C }{ \psi_{,c} } \partdif{^2 C }{ p^{ab} \partial \pi }
        + 2 \partdif{ C }{ \psi_{,cd} } \partial_d \left( \partdif{^2 C }{ p^{ab} \partial \pi } \right)
        + \partdif{ C }{ \pi } \partdif{^2 C }{ \psi_{,c} \partial p^{ab} }
        - 2 \partdif{ C }{ \pi } \partial_d \left( \partdif{^2 C }{ \psi_{,cd} \partial p^{ab} } \right)
        - \partdif{ ( \beta D^c ) }{ p^{ab} }
        - \partial_d \left( \beta \partdif{ D^d }{ p^{ab}_{,c} } \right)
    \right\}
\\ &
    + \partial_{cd} \theta^{ab} \left\{ 
        \partdif{ C }{ q_{ef,cd} } \partdif{^2 C }{ p^{ab} \partial p^{ef} }
        - \partdif{ C }{ p^{ef} } \partdif{^2 C }{ q_{ef,cd} \partial p^{ab} }
        + \partdif{ C }{ \psi_{,cd} } \partdif{^2 C }{ p^{ab} \partial \pi }
        - \partdif{ C }{ \pi } \partdif{^2 C }{ \psi_{,cd} \partial p^{ab} }
        - \beta \partdif{ D^c }{ p^{ab}_{,d} }
    \right\}.
\end{split}
    \label{eq:allst_dist-eqn-sol-metric}
\end{equation}
We disregard the zeroth order derivative of $\theta^{ab}$ because it does not provide useful information.

Before we can attempt to interpret this equation, we must first separate out all the different tensor combinations.  Since $\theta^{ab}$ is arbitrary, the coefficients of each unique tensor combination must vanish independently.
For example, suppose that ${0=B_{ab}\theta^{ab}}$. If $B_{ab}$ can be decomposed in terms of $q_{ab}$ and $p_\T^{ab}$, we find,
\begin{equation}
    0 = q_{ab} \theta^{ab} B_0 + p^\T_{ab} \theta^{ab} B_1 + p^\T_{ac} p^\T_{bd} q^{cd} \theta^{ab} B_2 + p^\T_{ac} p^\T_{bd} p_\T^{cd} \theta^{ab} B_3 + \ldots
\end{equation}
For this to be satisfied for general metrics, each coefficient $B_I$ must vanish independently.

When we substitute ${C=C(q,p,\bp,R,\psi,\pi,\Delta,\gamma)}$ into \eqref{eq:allst_dist-eqn-sol-metric}, there are many complicated tensor combinations that need to be considered, so for convenience we define
${X_a:=q^{bc}\partial_{a}q_{bc}}$.
We evaluate each term in the $\partial_{cd}\theta^{ab}$ bracket, and write them in \appref{sec:allst_extras}, in \eqref{eq:allst_extras_d2theta}.
So the linearly independent terms depending on $\partial_{cd}\theta^{ab}$ produce the following conditions,
\begin{subequations}
\begin{align} 
    \partial_{ab} \theta^{ab} : 0 & = 
        \partdif{ C }{ R } \partdif{ C }{ \bp } + \beta,
    \label{eq:allst_d2theta_1}
\\
\begin{split}
    q_{ab} \partial^2 \theta^{ab} : 0 & = 
        \partdif{ C }{ p } \partdif{^2 C }{ p \partial R }
        - \partdif{ C }{ R } \left(
            \partdif{^2 C }{ p^2 } 
            + \third \partdif{ C }{ \bp }
        \right)
        + \half \partdif{ C }{ \Delta } \partdif{^2 C }{ \pi \partial p }
        - \half \partdif{ C }{ \pi } \partdif{^2 C }{ p \partial \Delta },
\end{split}
    \label{eq:allst_d2theta_2} 
\\
    q_{ab} p_\T^{cd} \partial_{cd} \theta^{ab} : 0 & =
        \partdif{ C }{ R } \partdif{^2 C }{ p \partial \bp }
        - \partdif{ C }{ \bp } \partdif{^2 C }{ p \partial R },
    \label{eq:allst_d2theta_3}
\\
    p^\T_{ab} \partial^2 \theta^{ab} : 0 & =
        \partdif{ C }{ p } \partdif{^2 C }{ \bp \partial R }
        - \partdif{ C }{ R } \partdif{^2 C }{ p \partial \bp }
        + \half \partdif{ C }{ \Delta } \partdif{^2 C }{ \pi \partial \bp }
        - \half \partdif{ C }{ \pi } \partdif{^2 C }{ \bp \partial \Delta },
    \label{eq:allst_d2theta_4}
\\
    p^\T_{ab} p_\T^{cd} \partial_{cd} \theta^{ab} : 0 & =
        \partdif{ C }{ R } \partdif{^2 C }{ \bp^2 }
        - \partdif{ C }{ \bp } \partdif{^2 C }{ \bp \partial R }.
    \label{eq:allst_d2theta_5}
\end{align}
    \label{eq:allst_d2theta}%
\end{subequations}
We then evaluate each term in the ${\partial_c\theta^{ab}}$ bracket of \eqref{eq:allst_dist-eqn-sol-metric} and write them in \eqref{eq:allst_extras_dtheta}.
There are many unique terms which should be considered here, but in this case most of these are already solved by a constraint which satisfies \eqref{eq:allst_d2theta}.  
So the equations containing new information are,
\begin{subequations}
\begin{align}
    \partial_a \psi \partial_b \theta^{ab} : 0 & =
        \left( 
            2 \partdif{ C }{ R } \partial_\psi
            - \partdif{ C }{ \Delta }
        \right) \partdif{ C }{ \bp }
        + \partial_\psi \beta,
    \label{eq:allst_dtheta_1}
\\  
\begin{split}
    q_{ab} \partial^c \psi \partial_c \theta^{ab} : 0 & =
            \left( \half \partdif{ C }{ \Delta } - 4 \partdif{ C }{ R } \partial_\psi \right) \left( \partdif{^2 C }{ p^2 } + \third \partdif{ C }{ \bp } \right)
            + \half \partdif{ C }{ \Delta } \partdif{ C }{ \bp }
            + \partdif{ C }{ p } \left( \half \partdif{^2 C }{ p \partial \Delta } + 4 \partial_\psi \partdif{^2 C }{ p \partial R } \right)
\\ &
            + 2 \left( \partdif{ C }{ \gamma } + \partdif{ C }{ \Delta } \partial_\psi \right) \partdif{^2 C }{ \pi \partial p }
            + 2 \partdif{ C }{ \pi } \left( \partdif{^2 C }{ p \partial \gamma } - \partial_\psi \partdif{^2 C }{ p \partial \Delta } \right)
            - \pi \partdif{ \beta }{ p },
\end{split}
    \label{eq:allst_dtheta_2}
\\
\begin{split}
    p^\T_{ab} \partial^c \psi \partial_c \theta^{ab} : 0 & =
            \left( \half \partdif{ C }{ \Delta } - 4 \partdif{ C }{ R } \right) \partdif{^2 C }{ p \partial \bp }
            + \partdif{ C }{ p } \left( \half \partdif{^2 C }{ \bp \partial \Delta } + 4 \partial_\psi \partdif{^2 C }{ \bp \partial R } \right)
            + 2 \left( \partdif{ C }{ \gamma }
            + \partdif{ C }{ \Delta } \partial_\psi \right) \partdif{^2 C }{ \pi \partial \bp }
\\ &
            + 2 \partdif{ C }{ \pi } \left( \partdif{^2 C }{ \bp \partial \gamma } - \partial_\psi \partdif{^2 C }{ \bp \partial \Delta } \right)
            - \pi \partdif{ \beta }{ \bp },
\end{split}
    \label{eq:allst_dtheta_3}
%\end{align}
\\
%\begin{align}
    q_{ab} p_\T^{cd} \partial_d \psi \partial_c \theta^{ab} : 0 & =
        \left( 2 \partdif{ C }{ R } \partial_\psi - \partdif{ C }{ \Delta } \right) \partdif{^2 C }{ p \partial \bp }
        - \partdif{ C }{ \bp } \left( 2 \partial_\psi \partdif{^2 C }{ p \partial R } + \partdif{^2 C }{ p \partial \Delta } \right),
    \label{eq:allst_dtheta_4}
\\  
    p^\T_{ab} p_\T^{cd} \partial_d \psi \partial_c \theta^{ab} : 0 & =
        \left( 2 \partdif{ C }{ R } \partial_\psi - \partdif{ C }{ \Delta } \right) \partdif{^2 C }{ \bp^2 }
        - \partdif{ C }{ \bp } \left( 2 \partial_\psi \partdif{^2 C }{ \bp \partial R } + \partdif{^2 C }{ \bp \partial \Delta } \right),
    \label{eq:allst_dtheta_5}
\end{align}
\begin{align}
    X_a \partial_b \theta^{ab} : 0 & =
        \partdif{ C }{ R } \left( 1 + 2 \partial_q \right) \partdif{ C }{ \bp }
        + \partial_q \beta,
    \label{eq:allst_dtheta_6}
\\  
\begin{split}
    q_{ab} X^c\partial_c \theta^{ab} : 0 & =
        \partdif{ C }{ p } \left( 4 \partial_q - 1 \right) \partdif{^2 C }{ p \partial R }
        - \partdif{ C }{ R } \left( 4 \partial_q + 1 \right) \left( \partdif{^2 C }{ p^2 } + \third \partdif{ C }{ \bp } \right)
        + \half \partdif{ C }{ \pi } \left( 1 - 4 \partial_q \right) \partdif{^2 C }{ \Delta \partial p }
\\ &
        + \half \partdif{ C }{ \Delta } \left( 1 + 4 \partial_q \right) \partdif{^2 C }{ \pi \partial p }
        - \third p \partdif{ \beta }{ p },
\end{split}
    \label{eq:allst_dtheta_7}
\\
\begin{split}
    p^\T_{ab} X^c \partial_c \theta^{ab} : 0 & =
        \partdif{ C }{ p } \left( 4 \partial_q - 1 \right) \partdif{^2 C }{ \bp \partial R }
        - \partdif{ C }{ R } \left( 4 \partial_q + 1 \right) \partdif{^2 C }{ \bp \partial p }
        + \half \partdif{ C }{ \pi } \left( 1 - 4 \partial_q \right) \partdif{^2 C }{ \bp \partial \Delta }
\\ &
        + \half \partdif{ C }{ \Delta } \left( 1 + 4 \partial_q \right) \partdif{^2 C }{ \bp \partial \pi }
        - \third p \partdif{ \beta }{ \bp },
\end{split}
    \label{eq:allst_dtheta_8}
%\end{align}
\\
%\begin{align}
    q_{ab} p_\T^{cd} X_d \partial_c \theta^{ab} : 0 & =
        \partdif{ C }{ R } \left( 1 + 2 \partial_q \right) \partdif{^2 C }{ p \partial \bp }
        + \partdif{ C }{ \bp } \left( 1 - 2 \partial_q \right) \partdif{^2 C }{ p \partial R },
    \label{eq:allst_dtheta_9}
\\  
    p^\T_{ab} p_\T^{cd} X_d \partial_c \theta^{ab} : 0 & =
        \partdif{ C }{ R } \left( 1 + 2 \partial_q \right) \partdif{^2 C }{ \bp^2 }
        + \partdif{ C }{ \bp } \left( 1 - 2 \partial_q \right) \partdif{^2 C }{ \bp \partial R },
    \label{eq:allst_dtheta_10}
\end{align}
\begin{align}
    \partial_a F \partial_b \theta^{ab} : 0 & =
        2 \partdif{ C }{ R } \partdif{^2 C }{ F \partial \bp } + \partdif{ \beta }{ F },
    \label{eq:allst_dtheta_11}
\\  
\begin{split}
    q_{ab} \partial^c F \partial_c \theta^{ab} : 0 & =
        2 \partdif{ C }{ p } \partdif{^3 C }{ F \partial p \partial R }
        - 2 \partdif{ C }{ R } \partdif{}{F} \left( \partdif{^2 C }{ p^2 } + \third \partdif{ C }{ \bp } \right)
        + \partdif{ C }{ \Delta } \partdif{^3 C }{ F \partial p \partial \pi }
        - \partdif{ C }{ \pi } \partdif{^3 C }{ F \partial p \partial \Delta }
        + \third \delta_{\!F}^{p} \partdif{ \beta }{ p },
\end{split}
    \label{eq:allst_dtheta_12}
%\end{align}
\\
%\begin{align}
\begin{split}
    p^\T_{ab} \partial^c F \partial_c \theta^{ab} : 0 & =
        2 \partdif{ C }{ p } \partdif{^3 C }{ F \partial \bp \partial R }
        - 2 \partdif{ C }{ R } \partdif{^3 C }{ F \partial p \partial \bp }
        + \partdif{ C }{ \Delta } \partdif{^3 C }{ F \partial \bp \partial \pi }
        - \partdif{ C }{ \pi } \partdif{^3 C }{ F \partial \bp \partial \Delta }
        + \third \delta_{\!F}^{p} \partdif{ \beta }{ \bp },
\end{split}
    \label{eq:allst_dtheta_13}
\\
    q_{ab} p_\T^{cd} \partial_d F \partial_c \theta^{ab} : 0 & = 
        \partdif{ C }{ R } \partdif{^3 C }{ F \partial p \partial \bp }
        - \partdif{ C }{ \bp } \partdif{^3 C }{ F \partial p \partial R },
    \label{eq:allst_dtheta_14}
\\
    p^\T_{ab} p_\T^{cd} \partial_c F \theta^{ab} : 0 & =
        \partdif{ C }{ R } \partdif{^3 C }{ F \partial \bp^2 }
        - \partdif{ C }{ \bp } \partdif{^3 C }{ F \partial \bp \partial R },
    \label{eq:allst_dtheta_15}
\end{align}%
    \label{eq:allst_dtheta}%
\end{subequations}
where $F\in\{p,\bp,R,\Delta,\gamma\}$.
These conditions strongly restrict the form of the constraint, but before we attempt to consolidate them we must find the conditions coming from the scalar field.

%%%%%%%%%%%%%%%%%%%%%%%%%%%%%%%%%%%%%%%%%%%%%%%%%%%%%%%%%%%%%%%%%%%%%

\subsection{\texorpdfstring{$\pi$}{Scalar momentum} route}
\label{sec:dist-eqn-sol-scalar}

Similar to the calculation using the metric momentum, we return to the distribution equation \eqref{eq:allst_dist-eqn} and take the functional derivative with respect to $\pi(z)$,
\begin{equation}
\begin{split}
    0 & =
    \funcdif{ C (x) }{ q_{ab} (y) } \left. \partdif{^2 C }{ \pi \partial p^{ab} } \right|_y \delta ( z, y )
    + \funcdif{ \partial C (x) }{ q_{ab} (y) \partial \pi (x) } \left. \partdif{ C }{ p^{ab} } \right|_y \delta ( z, x )
    + \funcdif{ C (x) }{ \psi (y) } \left. \partdif{^2 C }{ \pi^2 } \right|_y \delta ( z, y )
\\ &
    + \funcdif{ \partial C (x) }{ \psi (y) \partial \pi (x) } \left. \partdif{ C }{ \pi } \right|_y \delta ( z, x )
    - \delta ( z, x ) \left( \partdif{ ( \beta D^a ) }{ \pi } \partial_a \right)_x \delta ( x, y )
    - \xty,
\end{split}
\end{equation}
which can be rewritten as,
\begin{equation}
    0 = A ( x, y ) \delta ( z, y ) - A ( y, x ) \delta ( z, x ),
    \label{eq:allst_Adist}
\end{equation}
where,
\begin{equation}
\begin{split}
    A ( x, y ) & =
    \funcdif{ C (x) }{ q_{ab} (y) } \left. \partdif{^2 C }{ \pi \partial p^{ab} } \right|_y
    - \funcdif{ \partial C (y) }{ q_{ab} (x) \partial \pi (y) } \left. \partdif{ C }{ p^{ab} } \right|_x
    + \funcdif{ C (x) }{ \psi (y) } \left. \partdif{^2 C }{ \pi^2 } \right|_y
\\ &
    - \funcdif{ \partial C (y) }{ \psi (x) \partial \pi (y) } \left. \partdif{ C }{ \pi } \right|_x
    + \left( \partdif{ ( \beta D^a ) }{ \pi } \partial_a \right)_y \delta ( y, x ).
\end{split}
\end{equation}
Integrate \eqref{eq:allst_Adist} over $y$ to find $0=A(x,z)-A(x)\delta(x,z)$.  Multiply this by a test scalar $\eta(z)$ and integrate over $z$,
\begin{equation}
\begin{split}
    0 & = 
    \eta \left( \cdots \right)
    + \partial_a \eta \left\{ 
        \partdif{ C }{ q_{cd,a} } \partdif{^2 C }{ \pi \partial p^{cd} }
        + 2 \partdif{ C }{ q_{cd,ab} } \partial_b \left( \partdif{^2 C }{ \pi \partial p^{cd} } \right)
        + \partdif{ C }{ p^{cd} } \partdif{^2 C }{ q_{cd,a} \partial \pi }
        - 2 \partdif{ C }{ p^{cd} } \partdif{^2 C }{ q_{cd,ab} \partial \pi }
\right. \\ & \left.
        + \partdif{ C }{ \psi_{,a} } \partdif{^2 C }{ \pi^2 }
        + 2 \partdif{ C }{ \psi_{,ab} } \partial_b \left( \partdif{^2 C }{ \pi^2 } \right)
        + \partdif{ C }{ \pi } \partdif{^2 C }{ \psi_{,a} \partial \pi}
        - 2 \partdif{ C }{ \pi } \partial_b \left( \partdif{^2 C }{ \psi_{,ab} \partial \pi } \right)
        - \partdif{ ( \beta D^a ) }{ \pi }
    \right\}
\\ & 
    + \partial_{ab} \eta \left\{
        \partdif{ C }{ q_{cd,ab} } \partdif{^2 C }{ \pi \partial p^{cd} }
        - \partdif{ C }{ p^{cd} } \partdif{^2 C }{ q_{cd,ab} \partial \pi }
        + \partdif{ C }{ \psi_{,ab} } \partdif{^2 C }{ \pi^2 }
        - \partdif{ C }{ \pi } \partdif{^2 C }{ \psi_{,ab} \partial \pi }
    \right\}.
\end{split}
    \label{eq:allst_dist-eqn-sol-scalar}
\end{equation}
We evaluate each of the terms for $\partial_{ab}\eta$, and write them in \eqref{eq:allst_extras_d2eta}.
From these, we find the independent equations,
\begin{subequations}
\begin{align}
    \partial^2 \eta : 0 & =
        \partdif{ C }{ p } \partdif{^2 C }{ \pi \partial R }
        - \partdif{ C }{ R } \partdif{^2 C }{ \pi \partial p }
        + \half \partdif{ C }{ \Delta } \partdif{^2 C }{ \pi^2 }
        - \half \partdif{ C }{ \pi } \partdif{^2 C }{ \Delta \partial \pi },
    \label{eq:allst_d2eta_1}
\\
    p_\T^{ab} \partial_{ab} \eta : 0 & =
        \partdif{ C }{ R } \partdif{^2 C }{ \pi \partial \bp }
        - \partdif{ C }{ \bp } \partdif{^2 C }{ \pi \partial R }.
    \label{eq:allst_d2eta_2}
\end{align}
    \label{eq:allst_d2eta}%
\end{subequations}
Then, evaluating all the terms for $\partial_a\eta$, and writing them in \eqref{eq:allst_extras_deta}.
Therefore, ignoring terms solved by \eqref{eq:allst_d2eta}, the equations we get from $\partial_a\eta$ are,
\begin{subequations}
\begin{align}
\begin{split}
    \partial^a \psi \partial_a \eta : 0 & =
        \left( 
            \half \partdif{ C }{ \Delta}
            - 4 \partdif{ C }{ R } \partial_\psi
        \right) \partdif{^2 C }{ \pi \partial p }
        + \partdif{ C }{ p } \left(
            \half \partdif{^2 C }{ \Delta \partial \pi } + 4 \partial_\psi \partdif{^2 C }{ R \partial \pi }
        \right)
        + 2 \left(
            \partdif{ C }{ \gamma }
            + \partdif{ C }{ \Delta } \partial_\psi
        \right) \partdif{^2 C }{ \pi^2 }
\\ &
        + 2 \partdif{ C }{ \pi } \left( 
            \partdif{^2 C }{ \gamma \partial \pi }
            - \partial_\psi \partdif{^2 C }{ \Delta \partial \pi }
        \right)
        - \left( \beta + \pi \partdif{ \beta }{ \pi }\right),
\end{split}
    \label{eq:allst_deta_1}
\\
    p_\T^{ab} \partial_b \psi \partial_a \eta : 0 & =
        \left(
            \partdif{ C }{ R } \partial_\psi
            - \half \partdif{ C }{ \Delta }
        \right) \partdif{^2 C }{ \pi \partial \bp }
        - \partdif{ C }{ \bp }
        \left(
            \partial_\psi \partdif{^2 C }{ \pi \partial R }
            + \half \partdif{^2 C }{  \pi \partial \Delta }
        \right),
    \label{eq:allst_deta_2}
\end{align}
\begin{align}
\begin{split}
    X^a \partial_a \eta : 0 & =
        \partdif{ C }{ p } \left( 4 \partial_q - 1 \right) \partdif{^2 C }{ \pi \partial R }
        - \partdif{ C }{ R } \left( 4 \partial_q + 1 \right) \partdif{^2 C }{ \pi \partial p }
        + \half \partdif{ C }{ \Delta } \left( 1 + 4 \partial_q \right) \partdif{^2 C }{ \pi^2 }
\\ &
        + \half \partdif{ C }{ \pi } \left( 1 - 4 \partial_q \right) \partdif{^2 C }{ \pi \partial \Delta }
        - \third p \partdif{ \beta }{ \pi },
\end{split}
    \label{eq:allst_deta_3}
\\
    p_\T^{ab} X_b \partial_a \eta : 0 & =
        \partdif{ C }{ R } \left( 1 + 2 \partial_q \right) \partdif{^2 C }{ \pi \partial \bp }
        + \partdif{ C }{ \bp } \left( 1 - 2 \partial_q \right) \partdif{^2 C }{ \pi \partial R },
    \label{eq:allst_deta_4}
\end{align}
\begin{align}
\begin{split}
    \partial^a F \partial_a \eta : 0 & =
        \partdif{ C }{ p } \partdif{^3 C }{ F \partial \pi \partial R }
        - \partdif{ C }{ R } \partdif{^3 C }{ F \partial \pi \partial R }
        + \half \partdif{ C }{ \Delta } \partdif{^3 C }{ F \partial \pi^2 }
        - \half \partdif{ C }{ \pi } \partdif{^3 C }{ F \partial \pi \partial \Delta }
        + \frac{1}{6} \delta_{\!F}^{p} \partdif{ \beta }{ \pi },
\end{split}
    \label{eq:allst_deta_5}
\\
    p_\T^{ab} \partial_b F \partial_a \eta : 0 & =
        \partdif{ C }{ R } \partdif{^3 C }{ F \partial \pi \partial \bp }
        - \partdif{ C }{ \bp } \partdif{^3 C }{ F \partial \pi \partial R },
    \label{eq:allst_deta_6}
\end{align}
    \label{eq:allst_deta}%
\end{subequations}
where $F \in \{ p, \bp, R, \Delta, \gamma \}$.
Now that we have all of the conditions restricting the form of the Hamiltonian constraint, we can move on to consolidating and interpreting them.

%%%%%%%%%%%%%%%%%%%%%%%%%%%%%%%%%%%%%%%%%%%%%%%%%%%%%%%%%%

\section{Deriving the Hamiltonian constraint}
\label{sec:allst}

Now we have the full list of equations, we seek to find the restrictions they impose on the form of $C$.  Firstly, we use the condition from $\partial_{ab}\theta^{ab}$, \eqref{eq:allst_d2theta_1} to find
\begin{equation}
    \partdif{ C }{ R } = - \beta \left( \partdif{ C }{ \bp } \right)^{-1},
    \label{eq:allst_solving_1}
\end{equation}
which we substitute into the equation from ${p^\T_{ab}p_\T^{cd}\partial_{cd}\theta^{ab}}$, \eqref{eq:allst_d2theta_5},
\begin{equation}
\begin{split}
    0 & = 
    \partdif{ C }{ R } \partdif{^2 C }{ \bp^2 }
    - \partdif{ C }{ \bp } \partdif{^2 C }{ \bp \partial R }
\\ &
    = - 2 \beta \left( \partdif{ C }{ \bp } \right)^{-1} \partdif{^2 C }{ \bp^2 }
    + \partdif{ \beta }{ \bp }
\\ &
    = \beta \partdif{}{ \bp } \log \left\{ \beta \left( \partdif{ C }{ \bp } \right)^{-2} \right\},
\end{split}
    \label{eq:allst_solving_2}
\end{equation}
and because $\beta\to1$ in the classical limit and so cannot vanish generally, we find that,
\begin{equation}
    \beta = b_1 \, \left( \partdif{ C }{ \bp } \right)^2,
        \quad \mathrm{where} \;
    \partdif{ b_1 }{ \bp } = 0.
    \label{eq:allst_solving_3}
\end{equation}
Substituting this back into \eqref{eq:allst_solving_1} gives us
$\textstyle{\partdif{C}{R}=-b_1\partdif{C}{\bp}}$, and from this we can find the first restriction on the form of the constraint,
\begin{equation}
    C ( q, p, \bp, R, \psi, \pi, \Delta, \gamma ) = 
    C_1 (q, p, \psi, \pi, \Delta, \gamma, \chi_1 ),
\quad
    \mathrm{where} \;
    \chi_1 := \bp - \int_0^R b_1 ( q, p, x, \psi, \pi, \Delta, \psi ) \mathrm{d} x.
    \label{eq:allst_solving_4}
\end{equation}
Substituting this into the condition from $\partial_aF\partial_b\theta^{ab}$, \eqref{eq:allst_dtheta_11}, gives
\begin{equation}
    0 = \partdif{ b_1 }{ F } \left( \partdif{ C_1 }{ \chi_1 } \right)^2,
    \quad 
    \mathrm{for} \;
    F \in \{ p, \bp, R, \pi, \Delta, \gamma \},
    \label{eq:allst_solving_5}
\end{equation}
and therefore $b_1$ must only be a function of $q$ and $\psi$. Substituting this into \eqref{eq:allst_solving_4} leads to ${\chi_1=\bp-b_1R}$.
Turning to $X_a\partial_b\theta^{ab}$, \eqref{eq:allst_dtheta_6}, we find
\begin{equation}
    0 = \left( \partdif{ C_1 }{ \chi_1 } \right)^2 \left( \partial_q - 1 \right) b_1 ,
    \label{eq:allst_solving_6}
\end{equation}
which is solved by ${b_1(q,\psi)=q\,b_2(\psi)}$.  This is as expected because it means both terms in $\chi_1$ have a density weight of two.
From this we see that ${\partial_a\psi\partial_b\theta^{ab}}$, \eqref{eq:allst_dtheta_1}, gives
\begin{equation}
    0 = \partdif{ C_1 }{ \chi_1 } \left( 
    q b_2' \partdif{ C_1 }{ \chi_1 }
    - \partdif{ C_1 }{ \Delta }
    \right)
    \label{eq:allst_solving_7}
\end{equation}
which provides further restrictions on the form of the constraint,
\begin{equation}
    C = C_2 ( q, p, \psi, \pi, \gamma, \chi_2 ),
    \quad
    \chi_2 := \bp - q \left( b_2 R - b_2' \Delta \right).
    \label{eq:allst_solving_8}
\end{equation}
Looking at $p^\T_{ab}\partial^2\theta^{ab}$, \eqref{eq:allst_d2theta_4},
\begin{equation}
\begin{split}
    0 & = \partdif{ C }{ p } \partdif{^2 C }{ \bp \partial R }
    - \partdif{ C }{ R } \partdif{^2 C }{ p \partial \bp }
    + \half \partdif{ C }{ \Delta } \partdif{^2 C }{ \pi \partial \bp }
    - \half \partdif{ C }{ \pi } \partdif{^2 C }{ \bp \partial \Delta },
\\ &
    = q b_2 \left\{
    - \partdif{ C_2 }{ p } \partdif{^2 C_2 }{ \chi_2^2 }
    + \partdif{ C_2 }{ \chi_2 } \partdif{^2 C }{ p \partial \chi_2 }
    + \frac{b_2'}{2b_2} \left(
        \partdif{ C_2 }{ \chi_2 } \partdif{^2 C_2 }{ \pi \partial \chi_2}
        - \partdif{ C_2 }{ \pi } \partdif{^2 C_2 }{ \chi_2^2 }
    \right) \right\}
\\ &
    = q b_2 \partdif{ C_2 }{ \chi_2 } \left( \partdif{ C_2 }{ p } + \frac{ b_2' }{ 2 b_2 } \partdif{ C_2 }{ \pi } \right) \partdif{}{ \chi_2 } \log \left\{
     \left( \partdif{ C_2 }{ p } + \frac{ b_2' }{ 2 b_2 } \partdif{ C_2 }{ \pi } \right) \left( \partdif{ C_2 }{ \chi_2 } \right)^{-1}
    \right\},
\end{split}
    \label{eq:allst_solving_9}
\end{equation}
and because $b_2$ is a non-zero constant in the classical limit, this can be integrated to find
\begin{equation}
    \partdif{ C_2 }{ p } 
    + \frac{ b_2' }{ 2 b_2 } \partdif{ C_2 }{ \pi }
    = g_1 ( q, p, \psi, \pi, \gamma ) \partdif{ C_2 }{ \chi_2 },
    \label{eq:allst_solving_10}
\end{equation}
where $g_1$ is a unknown function arising as an integration constant, and needs to be determined.
This provides a further restriction on the form of the constraint,
\begin{equation}
\begin{gathered}
    C = C_3 \left( q, \psi, \gamma, \Pi, \chi_3 \right),
        \quad
    \Pi := \pi - \frac{ b_2' }{ 2 b_2 } p,
        \quad
    \chi_3 := P - q \left( b_2 R - b_2' \Delta \right)
    + \int_0^p g_1 \left( q, x, \psi, \Pi + \frac{ b_2' }{ 2 b_2 } x, \gamma \right) \mathrm{d} x.
\end{gathered}
    \label{eq:allst_solving_11}
\end{equation}
Substituting this into the equation from $\partial^2\eta$, \eqref{eq:allst_d2eta_1}, gives
\begin{equation}
    0 = q \, b_2 \left( \partdif{ C_3 }{ \chi_3 } \right)^2 \partdif{}{\pi} g_1 \left( q, p, \psi, \pi, \gamma \right),
    \label{eq:allst_solving_12}
\end{equation}
and therefore,
\begin{equation}
    \chi_3 = P - q \left( b_2 R - b_2' \Delta \right)
    + \int_0^p g_1 \left( q, x, \psi, \gamma \right) \mathrm{d} x.
    \label{eq:allst_solving_13}
\end{equation}
Evaluating the equation from $q_{ab}\partial^2\theta^{ab}$, \eqref{eq:allst_d2theta_2}, gives
\begin{equation}
    0 = \third q b_2 \left( \partdif{ C_3 }{ \chi_3 } \right)^2 \left( 1 + 3 \partdif{}{ p } \right) g_1 \left( q, p, \psi, \gamma \right),
    \label{eq:allst_solving_14}
\end{equation}
which can be integrated to find 
${g_1=g_2\left(q,\psi,\gamma\right)-p/3}$
and therefore \eqref{eq:allst_solving_13} becomes,
\begin{equation}
    \chi_3 = \bp - \frac{1}{6} p^2 + g_2 \, p - q \left( b_2 R - b_2' \Delta \right).
    \label{eq:allst_solving_15}
\end{equation}
Then look at ${p^\T_{ab}X^c\partial_c\theta^{ab}}$, \eqref{eq:allst_dtheta_8}, from which can be found
\begin{equation}
    0 = 2 q b_2 \partdif{ C_3 }{ \chi_3 } \partdif{^2 C_3 }{ \chi_3^2 } \left( 2 \partial_q - 1 \right) g_2,
    \label{eq:allst_solving_16}
\end{equation}
which can be solved by,
${g_2 \left( q, \psi, \gamma \right) = \sqrt{q} \, g_3 \! \left( \psi, \gamma \right)}$
if we assume that 
$\textstyle{\partdif{^2C_3}{\chi_3^2}\neq0}$ 
generally, which is true for any deformation dependent on curvature
$\textstyle{\partdif{\beta}{\chi_3}\neq0}$.
This is what is expected for the density weight of each term in $\chi_3$ to match.

We now look at 
${p^\T_{ab}\partial^c\gamma\partial_c\theta^{ab}}$, 
which is \eqref{eq:allst_dtheta_13} with $F=\gamma$, 
\begin{equation}
    0 = q^{3/2} b_2 \partdif{ g_3 }{ \gamma } \partdif{ C_3 }{ \chi_3 } \partdif{^2 C_3 }{ \chi_3^2 },
    \label{eq:allst_solving_17}
\end{equation}
which is true when 
${g_3 = g_3 \left( \psi \right)}$.

At this point it gets harder to progress further as we have done so far.  To review, we have restricted the constraint and deformation to the forms,
\begin{equation}
\begin{gathered}
    C \left( q, p, \bp, R, \psi, \pi, \Delta, \gamma \right) = C_3 \left( q, \psi, \Pi, \gamma, \chi_3 \right),
        \quad
    \beta = q \, b_2 \left( \psi \right) \left( \partdif{ C_3 }{ \chi_3 } \right)^2,
        \\
    \Pi = \pi - \frac{b_2'}{2b_2} p,
        \quad
    \chi_3 = \bp - \frac{1}{6} p^2 + p \sqrt{q} \, g_3 \left( \psi \right) - q \left( b_2 R - b_2' \Delta \right),
\end{gathered}
    \label{eq:allst_solving_review}
\end{equation}
which satisfy all the conditions in \eqref{eq:allst_d2theta}, \eqref{eq:allst_d2eta}, \eqref{eq:allst_deta} and \eqref{eq:allst_dtheta} apart from the conditions for $q_{ab}\partial^c\psi\partial_c\theta^{ab}$, \eqref{eq:allst_dtheta_2}, and  $\partial^a\psi\partial_a\eta$, \eqref{eq:allst_deta_1}.  
As it stands, these are not easy to solve.

%%%%%%%%%%%%%%%%%%%%%%%%%%%%%%%%%%%%%%%%%%%%%%%%%%%%%%%%%%%

\subsection{Solving the fourth order constraint to advise the general case}
\label{sec:allst_fourth}

To break this impasse, we use a test ansatz for the constraint which contains up to four orders in momenta,
\begin{equation}
    C_3 \to C_0 + C_{(\Pi)} \Pi + C_{(\Pi^2)} \Pi^2 + C_{(\Pi^3)} \Pi^3 + C_{(\Pi^4)} \Pi^4
    + C_{(\chi)} \chi_3 + C_{(\chi^2)} \chi_3^2 + C_{(\Pi\chi)} \Pi \chi_3 + C_{(\Pi^2\chi)} \Pi^2 \chi_3,
\end{equation}
where each coefficient is an unknown function to be determined dependent on $q$, $\psi$ and $\gamma$.  There is an asymmetric term included in $\chi_3$ determined by the function $g_3\left(\psi\right)$, so we do not restrict ourselves to only even orders of momenta, unlike ref.~\cite{cuttell2018}.

Substituting this into \eqref{eq:allst_deta_1}, we can separate out the multiplier of each unique combination of variables as an independent equation.  For each of the terms which are the multipliers of 5 or 6 orders of momenta, we find a condition specifying that the constraint coefficients for terms 3 or 4 orders of momenta must not depend on $\gamma$, e.g. 
${\textstyle{\partdif{}{\gamma}C_{(\chi^2)}=0}}$, 
${\textstyle{\partdif{}{\gamma}C_{(\Pi^3)}=0}}$.  
Since $\gamma$ depends on two spatial derivatives, we see that each term in the constraint must not depend on a higher order of spatial derivatives than it does momenta.  Indeed, if we include higher powers of spatial derivatives in the ansatz, we quickly find them ruled out in a similar fashion.  Therefore, we use this information to further expand our ansatz,
\begin{equation}
\begin{split}
    C_3 & \to C_\0 + C_{(\gamma)} \gamma + C_{(\gamma^2)} \gamma^2 + C_{(\Pi)} \Pi + C_{(\Pi\gamma)} \Pi \gamma + C_{(\Pi^2)} \Pi^2 
    + C_{(\Pi^2\gamma)} \Pi^2 \gamma + C_{(\Pi^3)} \Pi^3 + C_{(\Pi^4)} \Pi^4
\\ &
    + C_{(\chi)} \chi_3 + C_{(\chi\gamma)} \chi_3 \gamma + C_{(\chi^2)} \chi_3^2
    + C_{(\Pi\chi)} \Pi \chi_3 + C_{(\Pi^2\chi)} \Pi^2 \chi_3,
\end{split}
\end{equation}
where each coefficient is now a function of $q$ and $\psi$.

One can find all the necessary conditions from \eqref{eq:allst_deta_1}, for which the solution also satisfies \eqref{eq:allst_dtheta_2}.  We will show a route which can taken to progressively restrict $C$.  The condition coming from $\bp^2$ is solved by
\begin{equation}
    C_{(\Pi^2\chi)} = \half C_{(\chi^2)} \left( \frac{ C_{(\chi\gamma)} }{ 2 q b_2 C_{(\chi^2)} } + \frac{ 7 b_2^{\prime2} }{ 8 b_2^2 } - \frac{ b_2'' }{ b_2 } \right)^{-1}.
\end{equation}
Looking at $\gamma^2$, we find,
\begin{equation}
    C_{(\Pi^2\gamma)} = \quarter C_{(\chi\gamma)} \left( \frac{ 2 C_{(\gamma^2)} }{ q b_2 C_{(\chi\gamma)} } + \frac{ 7 b_2^{\prime2} }{ 8 b_2^2 } - \frac{ b_2'' }{ b_2 } \right)^{-1},
\end{equation}
and from $\gamma\bp$,
\begin{equation}
    C_{(\gamma^2)} = \frac{ C_{(\chi\gamma)}^2 }{ 4 C_{(\chi^2)} },
\end{equation}
and from $\pi^4$,
\begin{equation}
    C_{(\Pi^4)} = \frac{1}{16} C_{(\chi^2)} \left( \frac{ C_{(\chi\gamma)} }{ 2 q b_2 C_{(\chi^2)} } + \frac{ 7 b_2^{\prime2} }{ 8 b_2^2 } - \frac{ b_2'' }{ b_2 } \right)^{-1}.
\end{equation}
At this point all the other conditions coming from four momenta are already solved.  Turning to the third order, the condition from $\pi\bp$ is solved by,
\begin{equation}
\begin{split}
    C_{(\Pi^3)} & = \frac{1}{12} \left\{ 
    C_{(\Pi\chi)} \left[ \frac{1}{qb_2} 
        \left( \frac{ 3 C_{(\chi\gamma)} }{ 2 C_{(\chi^2)} } - \frac{ 2 C_{(\Pi\gamma)} }{ C_{(\Pi\chi)} } \right) + \frac{ 7 b_2^{\prime2} }{ 8 b_2^2 } - \frac{ b_2'' }{ b_2 }
    \right]
        - \sqrt{q} \, C_{(\chi^2)} \left( 4 g_3' - \frac{3g_3b_2'}{b_2} \right) 
    \right\}
\\ &
    \times \left( \frac{ C_{(\chi\gamma)} }{ 2 q b_2 C_{(\chi^2)} } + \frac{ 7 b_2^{\prime2} }{ 8 b_2^2 } - \frac{ b_2'' }{ b_2 } \right)^{-2},
\end{split}
\end{equation}
and from $\pi\gamma$,
\begin{equation}
    C_{(\Pi\gamma)} = \frac{ C_{(\Pi\chi)} C_{(\chi\gamma)} }{ 2 C_{(\chi^2)} },
\end{equation}
and from $\pi^3$,
\begin{equation}
    C_{(\Pi\chi)} = \frac{-1}{2} \sqrt{q} \, C_{(\chi^2)} \left( 4 g_3' - \frac{ 3 g_3 b_2' }{ b_2 } \right) \left( \frac{ C_{(\chi\gamma)} }{ 2 q b_2 C_{(\chi^2)} } + \frac{ 7 b_2^{\prime2} }{ 8 b_2^2 } - \frac{ b_2'' }{ b_2 } \right)^{-1},
\end{equation}
which is all the new information from third order.  The only new condition coming from second order is solved by,
\begin{equation}
\begin{split}
    C_{(\Pi^2)} & = \left\{ 
    \quarter C_{(\Pi\chi)} \left[ \frac{1}{qb_2} 
        \left( \frac{ C_{(\chi\gamma)} }{ C_{(\chi^2)} } - \frac{ C_{(\gamma)} }{ C_{(\chi)} } \right) + \frac{ 7 b_2^{\prime2} }{ 8 b_2^2 } - \frac{ b_2'' }{ b_2 }
    \right]
        + \frac{q}{16} C_{(\chi^2)} \left( 4 g_3' - \frac{3g_3b_2'}{b_2} \right)^2 
    \right\}
\\ & \times
    \left( \frac{ C_{(\chi\gamma)} }{ 2 q b_2 C_{(\chi^2)} } + \frac{ 7 b_2^{\prime2} }{ 8 b_2^2 } - \frac{ b_2'' }{ b_2 } \right)^{-2},
\end{split}
\end{equation}
and the only new condition coming from first order is solved by,
\begin{equation}
\begin{split}
    C_{(\Pi)} & = \frac{-1}{4} \sqrt{q} \, C_{(\chi)} \left( 4 g_3' - \frac{ 3 g_3 b_2' }{ b_2 } \right)
    \left\{ \frac{1}{qb_2} \left( \frac{ C_{(\chi\gamma)} }{ C_{(\chi^2)} } - \frac{ C_{(\gamma)} }{ C_{(\chi)} } \right) + \frac{ 7 b_2^{\prime2} }{ 8 b_2^2 } - \frac{ b_2'' }{ b_2 }  \right\}
    \left( \frac{ C_{(\chi\gamma)} }{ 2 q b_2 C_{(\chi^2)} } + \frac{ 7 b_2^{\prime2} }{ 8 b_2^2 } - \frac{ b_2'' }{ b_2 } \right)^{-2},
\end{split}
\end{equation}
and likewise from the zeroth order,
\begin{equation}
    C_{(\chi\gamma)} = \frac{ 2 C_{(\gamma)} C_{(\chi^2)} }{ C_{(\chi)} }.
\end{equation}
When all of these terms are combined, we find the solution for the fourth order constraint,
\begin{equation}
    C = C_\0 
    + C_{(\chi)} \left( \chi_3 + \frac{ \Pi \left( \Pi - \Xi \right) }{ 4 \Omega } + \frac{ C_{(\gamma)} }{ C_{(\chi)} } \gamma \right)
    + C_{(\chi^2)} \left( \chi_3 + \frac{ \Pi \left( \Pi - \Xi \right) }{ 4 \Omega } + \frac{ C_{(\gamma)} }{ C_{(\chi)} } \gamma \right)^2,
        \label{eq:allst_constraint_4th}
\end{equation}
where
\begin{equation}
\begin{aligned}
    \Omega & =
    \frac{ C_{(\gamma)} }{ b_2 q C_{(\chi)} }
    + \frac{ 7 b_2^{\prime 2} }{ 8 b_2^2 }
    - \frac{ b_2'' }{ b _2 },
        &
    \Xi & = \sqrt{q} \left( 4 g_3' - \frac{ 3 g_3 b_2' }{ b_2 } \right).
\end{aligned}%
\end{equation}
If this solution is generalised to all orders,
\begin{equation}
    C = C_4 \left( q, \psi, \chi_4 \right),
        \quad
    \chi_4 = \bp - \frac{1}{6} p^2 + \sqrt{q} p g_3 - q \left( b_2 R - b_2' \Delta \right)
    + \frac{ \Pi \left( \Pi - \Xi \right) }{ 4 \Omega } 
    + \frac{ C_{(\gamma)} }{ C_{(\chi)} } \gamma,
\end{equation}
one can check that it satisfies all the conditions from \eqref{eq:allst_d2theta}, \eqref{eq:allst_dtheta}, \eqref{eq:allst_d2eta}, and \eqref{eq:allst_deta}.  It is possible that attempting to generalise from the fourth order constraint rather than continuing to work generally means that this is not the most general solution.  However, at least we now know a form of the constraint which \emph{can} solve all the conditions.

Now that we have a form for the general constraint, we seek to compare it to the low-curvature limit, when $C\to\chi_4C_\chi+C_\0$, and match terms with that found previously in ref.~\cite{cuttell2018}.  We find that,
\begin{equation}
    b_2 = \frac{ \sigma_\beta \omega_R^2 }{ 4 },
        \quad
    \sigma_\beta := \sgn{\beta} = \sgn{\beta_\0}.
        \quad
    C_\chi  = \frac{ 2 \sigma_\beta }{ \omega_R } \sqrt{ \frac{ \left| \beta_\0 \right| }{ q } },
        \quad
    C_\gamma  = \sqrt{ q \left| \beta_\0 \right| } \left( \frac{ \omega_\psi }{ 2 } + \omega_R'' \right),
\end{equation}
For convenience, we redefine the function determining the asymmetry, ${g_3=\xi/2}$,
and expand the constraint in terms of ${\R:=\chi_4/q}$, which has a density weight of zero.
This means that the general form of the deformed constraint is given by,
\begin{subequations}
\begin{gather}
    C = C \left( q, \psi, \R \right),
\quad
    \beta = \frac{ \sigma_\beta }{ q } \left( \partdif{ C }{ \R } \right)^2,
        \label{eq:allst_constraint-solution_differential} \\
\begin{split}
    \R & : = 
    \frac{ 2 \sigma_\beta }{ q \omega_R } \left( \bp - \frac{1}{6} p^2 \right)
    - \frac{ \omega_R }{ 2 } R
    + \omega_R' \Delta \psi
    + \left( \frac{\omega_\psi}{2} + \omega_R'' \right) \partial^a \psi \partial_a \psi
\\ &
    + \frac{ \sigma_\beta \omega_R }{ \omega_\psi \omega_R + \frac{ 3 }{ 2 } \omega_R^{\prime2} }
    \left\{ \frac{ 1 }{ 2 q } \left( \pi - \frac{ \omega_R' }{ \omega_R } p \right)^2 
    + \frac{ \xi }{ \sqrt{q} } \left(       
        \frac{ \omega_\psi }{ \omega_R } p 
        + \frac{ 3 \omega_R' }{ 2 \omega_R } \pi 
    \right)
    - \frac{ \xi' }{ \sqrt{q} } \left( \pi - \frac{ \omega_R' }{ \omega_R } p \right)
    \right\}.
        \label{eq:allst_constraint-solution_curvature}
\end{split}
\end{gather}%
    \label{eq:allst_constraint-solution}%
\end{subequations}
So all the momenta and spatial derivatives must combine in a specific way in the form of $\R$, which we might call the general kinetic term.

It is probably more appropriate to see the deformation function itself as the driver of deformations to the constraint, 
so we rearrange \eqref{eq:allst_constraint-solution_differential},
\begin{equation}
    \partdif{ C }{ \R } = \sqrt{ q \left| \beta \right| },
\end{equation}
which can be integrated to find,
\begin{equation}
    C = 
    \int_0^\R \sqrt{ q \left| \beta( q, \psi, r ) \right| } \, \mathrm{d} r 
    + C_\0 ( q, \psi ).
    \label{eq:allst_constraint-solution_integral}
\end{equation}
From either form of the general solution \eqref{eq:allst_constraint-solution_differential} or \eqref{eq:allst_constraint-solution_integral}, one can now understand the meaning of \eqref{eq:methodology_order_constraint}, which relates the highest orders of momenta in the constraint and the deformation, ${2n_C-n_\beta=4}$.  The differential form \eqref{eq:allst_constraint-solution_differential} is like
${n_\beta=2\left(n_C-2\right)}$,
and the integral form \eqref{eq:allst_constraint-solution_integral} is like
${n_C=2+n_\beta/2}$.

From the integral form of the solution \eqref{eq:allst_constraint-solution_integral}, we can now check a few examples of what constraint corresponds to certain deformations.  Here are a few examples of easily integrable functions with the appropriate classical limit,
\begin{equation}
\begin{split}
    \beta & = \beta_\0 \left( 1 + \beta_2 \R \right)^n
\\
    \to
    C & = C_\0 + 
\left\{
    \begin{aligned}
    & \frac{ 2 \sqrt{ q \left| \beta_\0 \right| } }{ \left(  n + 2 \right) \beta_2 }
    \left\{ \sgn{ 1 + \beta_2 \R } \left| 1 + \beta_2 \R \right|^{ \frac{ n + 2 }{ 2 } } - 1 \right\},
    & n \neq -2,
\\
    &
    \frac{ \sqrt{ q \left| \beta_\0 \right| } }{ \beta_2 } \sgn{ 1 + \beta_2 \R } \log \left| 1 + \beta_2 \R \right|,
    & n = -2,
    \end{aligned}
\right.
\\ &
    \simeq C_\0 + \sqrt{ q \left| \beta_\0 \right| } 
    \left\{ \R + \frac{ n \beta_2 }{ 4 } \R^2 + \cdots \right\}.
\end{split}
    \label{eq:allst_deftocon_linear}
\end{equation}
\begin{equation}
    \beta = \beta_\0 e^{\beta_2 \R}
    \; \to \;
    C = C_\0 
    + \frac{ 2 \sqrt{ q \left| \beta_\0 \right| } }{ \beta_2 } 
    \left( e^{ \beta_2 \R / 2} - 1 \right)
    \simeq
    C_\0 + \sqrt{ q \left| \beta_\0 \right| } \left( \R + \frac{ \beta_2 }{ 4 } \R^2 + \cdots \right),
        \label{eq:allst_deftocon_exp}
\end{equation}
\begin{equation}
    \beta = \beta_\0 \sech^2 \left( \beta_2 \R \right)
    \; \to \;
    C = C_\0 
    + \frac{ \sqrt{ q \left| \beta_\0 \right| } }{ \beta_2 } \gud \left( \beta_2 \R \right),
    \simeq
    C_\0 + \sqrt{ q \left| \beta_\0 \right| } \left( \R - \frac{ \beta_2^2 }{ 6 } \R^3 + \cdots \right),
     \label{eq:allst_deftocon_sech2}
\end{equation}
where $\textstyle{\gud(x):=\int_0^x\sech(t)\mathrm{d}t}$ is the Gudermannian function.
Most other deformation functions would need to be integrated numerically to find the constraint.
As can be seen from the small $\R$ expansions, it would be possible to constrain $\beta_\0$ and $\beta_2$ phenomenologically but the asymptotic behaviour of $\beta$ would be difficult to determine.

The simplest constraint that can be expressed as a polynomial of $\R$ that contains higher orders than the classical solution is given by,
\begin{equation}
    \beta = \beta_\0 \left( 1 + \beta_2 \R \right)^2
    \to
    C = C_\0 + \sqrt{ q \left| \beta_\0 \right| } \left( \R + \frac{ \beta_2 }{ 2 } \R^2 \right),
        \label{eq:allst_deftocon_simple}
\end{equation}
which is exactly what we found in \eqref{eq:allst_constraint_4th}.

%%%%%%%%%%%%%%%%%%%%%%%%%%%%%%%%%%%%%%%%%%%%%%%%%%%%%%%%%%

\subsection{Looking back at the constraint algebra}
\label{sec:allst_lookingback}

For this deformed constraint to mean anything, it must not reduce to the undeformed constraint through a simple transformation.  If we write the constraint as a function of the undeformed vacuum constraint $\bar{C}=\sqrt{q}\,\R$, we see that the deformation in the constraint algebra can be absorbed by a redefinition of the lapse functions,
\begin{subequations}
\begin{align}
    \{ C [ N ], C [ M ] \}
    & = \int \mathrm{d} x \mathrm{d} y N(x) M(y) \{ C (x), C (y) \},
\\ &
    = \int \mathrm{d} x \mathrm{d} y 
    \left( N \partdif{ C }{ \bar{C} } \right)_x
    \left( M \partdif{ C }{ \bar{C} } \right)_y \{ \bar{C} (x), \bar{C} (y) \},
\\ &
    = \int \mathrm{d} x \mathrm{d} y
    \left( \sigma_{\partial{}C} \bar{N} \right)_x \left( \sigma_{\partial{}C} \bar{M} \right)_y \{ \bar{C} (x), \bar{C} (y) \},
\\ &
    = \{ \bar{C} [ \sigma_{\partial{}C} \bar{N} ], \bar{C} [ \sigma_{\partial{}C} \bar{M} ] \},
\end{align}%
\end{subequations}
where 
$\bar{N}:=N\left|\partial{C}/\partial\bar{C}\right|$, 
$\bar{M}:=M\left|\partial{C}/\partial\bar{C}\right|$
and 
$\sigma_{\partial{}C}:=\sgn{\partial{C}/\partial{\bar{C}}}$, 
because the lapse functions should remain positive.
The other side of the equality,
\begin{equation}
\begin{split}
    D_a [ \beta q^{ab} ( N \partial_b M - \partial_b N M ) ],
    & =
    \int \mathrm{d} x D_a \beta q^{ab} ( N \partial_b M - \partial_b N M )
\\ &
    = \int \mathrm{d} x D_a \sigma_\beta \left( \partdif{ C }{ \bar{C} } \right)^2 ( N \partial_b M - \partial_b N M )
\\ &
    = \int \mathrm{d} x D_a \sigma_\beta ( \bar{N} \partial_b \bar{M} - \partial_b \bar{N} \bar{M} ),
\\ &
    = D_a [ \sigma_\beta q^{ab} ( \bar{N} \partial_b \bar{M} - \partial_b \bar{N} \bar{M} ) ],
\end{split}
\end{equation}
which we can combine to show the that the following two equations are equivalent,
\begin{subequations}
\begin{align}
    \{ C [ N ], C [ M ] \}
    & =
    D_a [ \beta q^{ab} ( N \partial_b M - \partial_b N M ) ],
\\
    \{ \bar{C} [ \sigma_{\partial{}C} \bar{N} ], \bar{C} [ \sigma_{\partial{}C} \bar{M} ] \}
    & =
    D_a [ \sigma_\beta q^{ab} ( \bar{N} \partial_b \bar{M} - \partial_b \bar{N} \bar{M} ) ].
\end{align}
\end{subequations}
The two $\sigma_{\partial{}C}$ on the left side should cancel out, but they are included here to show the limit to the redefinition of the lapse functions.  
While it may seem like we have regained the undeformed constraint algebra up to the sign $\sigma_\beta$ with a simple transformation, it shouldn't be taken to mean that this is actually the algebra of constraints.  That is, the above equation doesn't ensure that $\bar{C}\approx0$ instead of $C\approx0$ when on-shell.  The surfaces in phase space described by $\bar{C}=0$ and $C=0$ are different in general.

%%%%%%%%%%%%%%%%%%%%%%%%%%%%%%%%%%%%%%%%%%%%%%%%%%%%%%%%%%

\section{Cosmology}
\label{sec:cosmo}

We restrict to an isotropic and homogeneous space to find the background cosmological dynamics, following the definitions in \secref{sec:methodology_cosmo}.  Writing the constraint as ${C=C(a,\psi,\R)}$ where $\R=\R(a,\psi,\bar{p},\pi)$, the equations of motion are given by,
\begin{equation}
    \frac{ \dot{a} }{ N } 
    = \frac{ 1 }{ 6 a } \partdif{ \R }{ \bar{p} } \partdif{ C }{ \R },
        \quad
    \frac{ \dot{\bar{p}} }{ N } 
    = \frac{ - 1 }{ 6 a } \left( 
        \partdif{ C }{ a }
        + \partdif{ \R }{ a } \partdif{ C }{ \R }
    \right),
        \quad
    \frac{ \dot{\psi} }{ N }
    = \partdif{ \R }{ \pi } \partdif{ C }{ \R },
        \quad
    \frac{ \dot{\pi} }{ N }
    = - \partdif{ C }{ \psi }
        - \partdif{ \R }{ \psi } \partdif{ C }{ \R },
\end{equation}
into which we can substitute 
$\textstyle{\partdif{C}{\R}=a^3\sqrt{\left|\beta\right|}}$.
When we assume minimal coupling ($\omega_R'=\omega_\psi'=0$) and time-symmetry ($\xi=0$) the kinetic term is given by
\begin{equation}
    \R =
    \frac{ - 3 \sigma_\beta \bar{p}^2 }{ \omega_R a^2 }
    - \frac{ 3 k \omega_R }{ a^2 }
    + \frac{ \sigma_\beta \pi^2 }{ 2 \omega_\psi a^6 },
\end{equation}
and the equations of motion become,
\begin{equation}
    \frac{ \dot{a} }{ N } 
    = \frac{ - \sigma_\beta \bar{p} }{ \omega_R } \sqrt{ \left| \beta \right| },
\quad
    \frac{ \dot{\bar{p}} }{ N } 
    = \frac{ - 1 }{ 6 a } \partdif{ C }{ a }
    - a \sqrt{ \left| \beta \right| } \left(
        \frac{ \sigma_\beta \bar{p}^2 }{ \omega_R a^2 }
        + \frac{ k \omega_R }{ a^2 }
        - \frac{ \sigma_\beta \pi^2 }{ 2 \omega_\psi a^6}
    \right),
\quad
    \frac{ \dot{\psi} }{ N }
    = \frac{ \sigma_\beta \pi }{ \omega_\psi a^3 } \sqrt{ \left| \beta \right| },
\quad
    \frac{ \dot{\pi} }{ N }
    = - \partdif{ C }{ \psi }.
        \label{eq:cosmo_flrw_eom}
\end{equation}
To find the Friedmann equation, find the equation for 
$\textstyle{\H^2=\left(\frac{\dot{a}}{aN}\right)^2}$, 
and substitute in for $\R$,
\begin{equation}
    \H^2
    = \left| \beta \right| \frac{ \bar{p}^2 }{ \omega_R^2 }
    = \beta \left( \frac{ - \R }{ 3 \omega_R } - \frac{ k }{ a^2 } + \frac{ \sigma_\beta \pi^2 }{ 6 \omega_R \omega_\psi a^6 } \right),
        \label{eq:cosmo_friedmann_basic}
\end{equation}
and when the constraint is solved, $C\approx0$, then $\R$ can be found in terms of $C_\0$ .

\subsection{Cosmology with a perfect fluid}
\label{sec:cosmo_fluid}

\begin{figure}[t]
\begin{center}
	{\subfigure[$\beta=\beta_\0\left(1+\beta_2\R\right)^n$]{
		\label{fig:cosmo_friedmann_lin}
		\includegraphics[width=0.44\textwidth]{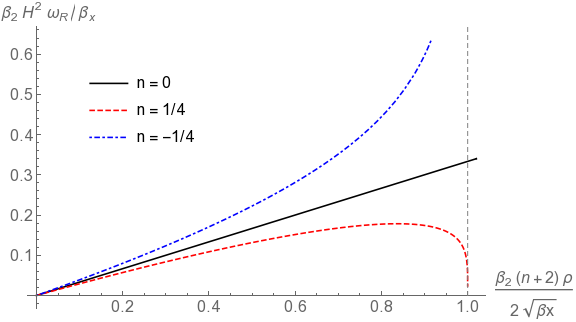}}}
	{\subfigure[$\beta\sim\exp{\R}$]{
		\label{fig:cosmo_friedmann_exp}
		\includegraphics[width=0.44\textwidth]{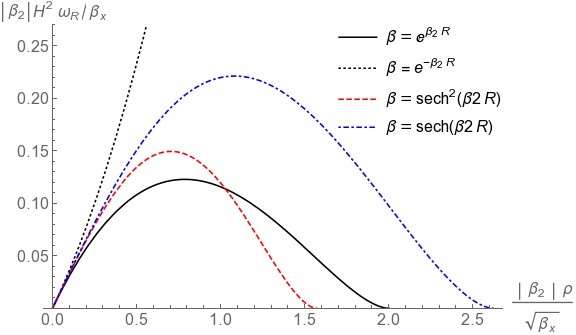}}}
\end{center}
    \caption[The Friedmann equation for various deformation functions $\beta(\R)$]
    {Behaviour of the Friedmann equation for various deformation functions $\beta(\R)$ when $k=0$ and $\beta_2>0$.}
    \label{fig:cosmo_friedmann}
\end{figure}

We here find the deformed Friedmann equations for various forms of the deformation.  For simplicity, we ignore the scalar field and include a perfect fluid $C_\0=a^3\rho(a)$.
From the deformation function $\beta=\beta_\0\left(1+\beta_2\R\right)^n$, solving the constraint \eqref{eq:allst_deftocon_linear} gives
\begin{equation}
    \R = 
    \left\{
    \begin{aligned}
    & \frac{ \sigma_2 }{ \beta_2 } \left\{ 
        \sigma_2 - \frac{ \left( n + 2 \right) \sigma_2 \beta_2 \rho }{ 2 \sqrt{ \left| \beta_\0 \right| } } 
    \right\}^{ \frac{ n + 2 }{ 2 } } 
    - \frac{ 1 }{ \beta_2 },
    & n \neq -2,
\\ &
    \frac{ \sigma_2 }{ \beta_2 } \exp \left( \frac{ - \sigma_2 \beta_2 \rho }{ \sqrt{ \left| \beta_\0 \right| } } \right)
    - \frac{ 1 }{ \beta_2 },
    & n = -2,
    \end{aligned}
    \right.
        \label{eq:cosmo_R_lin}
\end{equation}
where $\sigma_2:=\sgn{1+\beta_2\R}$.  When we simplify by assuming $\sigma_2=1$, the Friedmann equation is given by,
\begin{equation}
    \H^2 = \left\{
    \begin{aligned}
    & \left( \frac{ \beta_\0 }{ 3 \omega_R \beta_2 } \left[
        1 - \left( 
            1 - \frac{ \left( n + 2 \right) \beta_2 \rho }{ 2 \sqrt{ \left| \beta_\0 \right| } } 
        \right)^{\frac{2}{n+2}}
    \right] - \frac{ k \beta_\0 }{ a^2 } \right) \left( 
        1 - \frac{ \left( n + 2 \right) \beta_2 \rho }{ 2 \sqrt{ \left| \beta_\0 \right| } } 
    \right)^{\frac{2n}{n+2}},
    & n \neq -2,
\\&
    \left( \frac{ \beta_\0 }{ 3 \omega_R \beta_2 } \left[
        1 - \exp{ \left( \frac{ - \beta_2 \rho }{ \sqrt{ \left| \beta_\0 \right| } } \right) }
    \right] - \frac{ k \beta_\0 }{ a^2 } \right)
    \exp{ \left( \frac{ 2 \beta_2 \rho }{ \sqrt{ \left| \beta_\0 \right| } } \right) }
    & n = -2.
    \end{aligned}
    \right.
        \label{eq:cosmo_friedmann_lin}
\end{equation}
where there is a critical density, $\textstyle{\rho\to\frac{2\sqrt{\left|\beta_\0\right|}}{\left(n+2\right)\beta_2}}$, when $n\neq-2$.
To see the behaviour of the modified Friedmann equation for different values of $n$, look at \figref{fig:cosmo_friedmann_lin}.  For ${n>0}$, the Hubble rate vanishes as the universe approaches the critical energy density, this indicates that a collapsing universe reaches a turning point at which point the repulsive effect causes a bounce.  For ${0>n>-2}$, there appears a sudden singularity in $\H$ at finite $\rho$ (therefore finite $a$).
In the ${\rho\to\infty}$ limit, ${\H^2\sim{}e^{2\rho}}$ when ${n=-2}$ and ${\H^2\sim\rho^{\frac{2n}{n+2}}}$ when ${n<-2}$.

The singularities for $0>n>-2$ appear to be similar to sudden future singularities characterised in \cite{Cattoen:2005dx, FernandezJambrina:2006hj}.  However, the singularities here might instead be called sudden `past' singularities as they happen when $a$ is small (but non-zero) and $\rho$ is large.  Moreover, they happen for any perfect fluid with $w>-1$, i.e. including matter and radiation.

%%%%%%%%%%%%%%%%%%%%%%%%%%%%%%%%%%

For the deformation function 
${\beta=\beta_\0\exp{\left(\beta_2\R\right)}}$ 
from \eqref{eq:allst_deftocon_exp}, solving the constraint gives,
\begin{equation}
    \R = \frac{ 2 }{ \beta_2 } \log \left(
        1 - \frac{ \beta_2 \rho }{ 2 \sqrt{ \left| \beta_\0 \right| } }
    \right),
        \label{eq:cosmo_R_exp}
\end{equation}
and the Friedmann equation is given by,
\begin{equation}
    \H^2 = \left\{
        \frac{ - 2 \beta_\0 }{ 3 \omega_R \beta_2 } \log{ \left(
            1 - \frac{ \beta_2 \rho }{ 2 \sqrt{ \left| \beta_\0 \right| } }
        \right) } - \frac{ k \beta_2 }{ a^2 }
    \right\} \left(
        1 - \frac{ \beta_2 \rho }{ 2 \sqrt{ \left| \beta_\0 \right| } }
    \right)^2.
        \label{eq:cosmo_friedmann_exp}
\end{equation}
where there is a critical density,
$\textstyle{\rho\to\frac{2\sqrt{\left|\beta_\0\right|}}{\beta_2}}$.

For the deformation function
${\beta=\beta_\0\sech^2\left(\beta_2\R\right)}$ 
from \eqref{eq:allst_deftocon_sech2}, solving the constraint gives,
\begin{equation}
    \R = \frac{ - 1 }{ \beta_2 } \gud^{-1} \left( 
        \frac{ \beta_2 \rho }{ \sqrt{ \left| \beta_\0 \right| } } 
    \right).
        \label{eq:cosmo_R_sech2}
\end{equation}
Substituting this back into the deformation function gives,
\begin{equation}
    \beta = \beta_\0 \cos^2{ \left( 
        \frac{ \beta_2 \rho }{ \sqrt{ \left| \beta_\0 \right| } } 
    \right) }
        \label{eq:cosmo_def_sech2}
\end{equation}
and the Friedmann equation is given by
\begin{equation}
    \H^2 = \left\{
        \frac{ \beta_\0 }{ 3 \omega_R \beta_2 } \gud^{-1} \left( 
            \frac{ \beta_2 \rho }{ \sqrt{ \left| \beta_\0 \right| } } 
        \right) - \frac{ k \beta_\0 }{ a^2 }
    \right\} 
    \cos^2{ \left( 
        \frac{ \beta_2 \rho }{ \sqrt{ \left| \beta_\0 \right| } } 
    \right) }.
        \label{eq:cosmo_friedmann_sech2}
\end{equation}
where there is a critical density%
\footnote{where $\pi_{\circ}\approx 3.14$.},
$\textstyle{\rho\to\frac{\pi_\circ\sqrt{\left|\beta_\0\right|}}{2\left|\beta_2\right|}}$.  

These exponential-type deformation functions that we consider all predict a upper limit on energy density.  To illustrate this, we plot the modified Friedmann equations for these functions in \figref{fig:cosmo_friedmann_exp}.

%%%%%%%%%%%%%%%%%%%%%%%%%%%%%%%%%%%%%%%%%%%%%%%%%%%%%%%%%%%%

\subsection{Cosmology with a minimally coupled scalar field}
\label{sec:cosmo_scalar}

Since the metric and scalar kinetic terms must combine into one quantity, $\R$, a deformation function should not affect the relative structure between fields.  To illustrate this, take a free scalar field (without a potential) which is minimally coupled to gravity, and assume no perfect fluid component.  This means that the generalised potential term $C_\0$ will vanish, in which case solving the constraint, ${C\approx0}$, merely implies $\R=0$.  Consequently, since the deformation function $\beta$ is a function of $\R$, the only deformation remaining will be the zeroth order term ${\beta=\beta_\0\left(q,\psi\right)}$.  Combining the equations of motion \eqref{eq:cosmo_flrw_eom} allows us to find the resultant Friedmann equation,
\begin{equation}
    \H^2
    = \frac{ \omega_\psi \dot{\psi}^2 }{ 6 \omega_R N^2 }
    - \frac{ k \beta_\0 }{ a^2 },
\end{equation}
that is, the minimally-deformed case.
For $\beta\neq\beta_\0$, it is required that $\R$ must not vanish, which itself requires that $C_\0$ must be non-zero.  Therefore, for the dynamics to depend on a deformation which is a function of curvature, there must be a non-zero potential term which acts as a background against which the fields are deformed.

%%%%%%%%%%%%%%%%%%%%%%%%%%%%%%%%%%%%%%%%%%%%%%%%%5

\subsection{Deformation correspondence}
\label{sec:cosmo_deformation}

The form of the deformation used in the literature which includes holonomy effects is given by the cosine of the extrinsic curvature \cite{Cailleteau2012a, Mielczarek:2012pf, Cailleteau2013}.  Of particular importance to this is that the deformation vanishes and changes sign for high values of extrinsic curvature.  Since the extrinsic curvature is proportional to the Hubble expansion rate, write the deformation here as,
\begin{equation}
    \beta = \beta_\0 \cos \left( \beta_k \H \right).
        \label{eq:cosmo_beta_barrau}
\end{equation}
We wish to find $C(\R)$ and $\beta(\R)$ associated with this deformation. To do so, we need to find the relationship between the Hubble parameter ${\H=\dot{a}/aN}$ and the momentum $\bar{p}$, and thereby infer the form of $\beta(\R)$.  Then, using \eqref{eq:allst_constraint-solution_integral} we can find the constraint $C(\R)$.
So, using the equations of motion \eqref{eq:cosmo_flrw_eom}, we find 
\begin{equation}
    h = r \sqrt{\left|\cos{h}\right|},
\quad
    \mathrm{where, }
\quad
    h : = \beta_k \H,
\quad
    r : = \frac{ - \sigma_\beta \bar{p} }{ \omega_R a } \beta_k \sqrt{ \left| \beta_\0 \right| } ,
    \label{eq:cosmo_cosine_implicit}
\end{equation}
this is an equation which cannot be solved analytically for $h(r)$, and so must be solved numerically.

\begin{figure}[t]
	\begin{center}
	{\subfigure[$\beta\left(h\right)$]{
		\label{fig:cosmo_cosine_def-h}
		\includegraphics[width=0.40\textwidth]{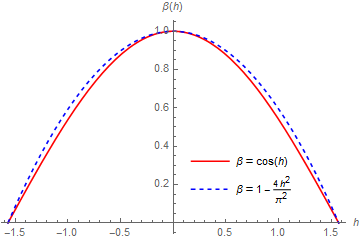}}}
	{\subfigure[$h(r)$]{
	    \label{fig:cosmo_cosine_def_matching}
		\includegraphics[width=0.40\textwidth]{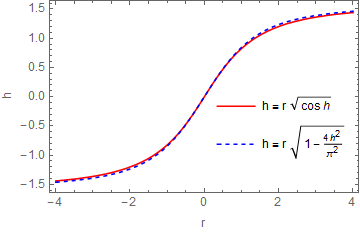}}}
	\\
	{\subfigure[$\beta\left(r\right)$]{
		\label{fig:cosmo_cosine_def-r}
		\includegraphics[width=0.40\textwidth]{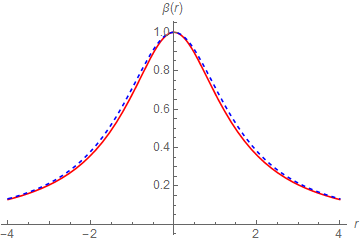}}}
	{\subfigure[$C_k\left(r\right)$]{
		\label{fig:cosmo_cosine_def_ck}
		\includegraphics[width=0.40\textwidth]{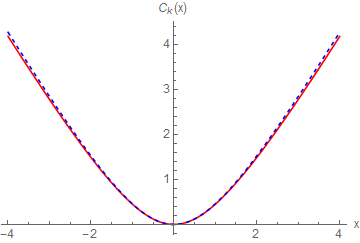}}}
	\end{center}
	\caption[Matching the Hubble expansion and canonical momentum for a cosine deformation]{
	Plot showing the process of starting from a deformation $\beta(h)$ \subref{fig:cosmo_cosine_def-h}, transforming $h(r)$ \subref{fig:cosmo_cosine_def_matching}, finding the new form of the deformation $\beta(r)$ \subref{fig:cosmo_cosine_def-r}, and finding the kinetic part of the constraint $C_k(r)$ \subref{fig:cosmo_cosine_def_ck}.  We include the function ${\beta=1-4\pi_\circ^{-2}h^{2}}$ (blue dashed line) because it has the same limits as ${\beta=\cos{h}}$ (red solid line) for the region ${|h|\leq\pi_{\circ}/2}$ but the transformation can be done analytically
	}
    	\label{fig:cosmo_cosine_def}
\end{figure}

For the general relation ${h=r\sqrt{|\beta(h)|}}$, there are similar $\beta$ functions which can be transformed analytically.  One example
is ${\beta(h)=1-4\pi_\circ^{-2}h^2}$,
which also has the same limits of ${\beta(0)=1}$ and ${\beta(h\to\pm\pi_{\circ}/2)=0}$,
and can be transformed to find
${\beta(r)=\left(1+4\pi_\circ^{-2}r^{2}\right)^{-1}}$.
In \figref{fig:cosmo_cosine_def}, we plot $\beta(h)$ and $h(r)$ in the region ${\left|h\right|\leq\pi_{\circ}/2}$.  After making the transformation, we find $\beta(r)$.
Note that, unlike for $h$, $\beta$ does not vanish for finite $r$.  So it seems that a deformation which vanishes for finite extrinsic curvature does not necessarily vanish for finite intrinsic curvature or metric momenta (at least not in the isotropic and homogeneous case).
In this respect, it matches the dynamics found for exponential-form deformations in \figref{fig:cosmo_friedmann}.

\begin{figure}[t]
	\begin{center}
	{\subfigure[$\beta\left(h\right)$]{
		\label{fig:cosmo_cosine_def2-h}
		\includegraphics[width=0.40\textwidth]{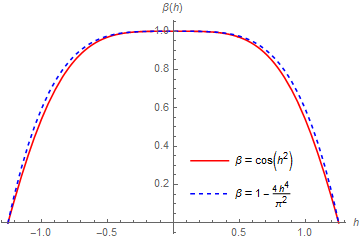}}}
	%
%	{\subfigure[$h(r)$]{
%	    \label{fig:cosmo_cosine_def2_matching}
%		\includegraphics[width=0.4\textwidth]{./def2_matching.png}}}
%	\\
	{\subfigure[$\beta\left(r\right)$]{
		\label{fig:cosmo_cosine_def2-r}
		\includegraphics[width=0.40\textwidth]{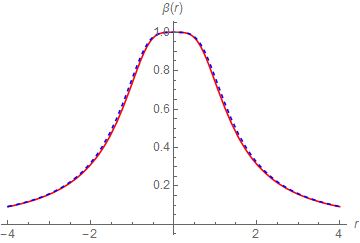}}}
	%
%	{\subfigure[$C_k\left(r\right)$]{
%		\label{fig:cosmo_cosine_def2_ck}
%		\includegraphics[width=0.4\textwidth]{./def2_ck.png}}}
	%
	\end{center}
	\caption[]{
	Plot showing transformations for the deformations given by ${\beta(h)=\cos{h^2}}$ (red solid line) and ${\beta(h)=1-4\pi_{\circ}^{-2}h^4}$ (blue dashed line).
	}
    	\label{fig:cosmo_cosine_def2}
\end{figure}

Returning to the solution for the constraint, \eqref{eq:allst_constraint-solution_integral}, reducing it to depending on only $a$ and $\bar{p}$ gives
\begin{equation}
    C = \frac{ - 6 a }{ \omega_R } \int_0^{\bar{p}} 
    \sigma_\beta \bar{p}'
    \sqrt{ \left| \beta \left( a, \bar{p}' \right) \right| } 
    \, \mathrm{d} \bar{p}'
    + C_\0 \left( a \right),
        \label{eq:cosmo_constraint-solution}
\end{equation}  
and transforming from $\bar{p}$ to $r$ as defined in \eqref{eq:cosmo_cosine_implicit}, while making the assumptions ${\sigma_\beta=1}$, ${N=1}$, ${\beta_\0=1}$, and ${\beta_k\sim\mathrm{constant}}$, this becomes
\begin{equation}
    C = \frac{ - 6 \omega_R a^3 }{ \beta_k^2 } C_k (r) + C_\0 (a),
\quad
    C_k (r) := \int_0^{r} r' \sqrt{ \beta ( r' ) } \, \mathrm{d} r'.
\end{equation}
We numerically integrate the solution for $\beta(r)$ found for when ${\beta=\cos(h)}$.  We plot the function $C_k(r)$ in \figref{fig:cosmo_cosine_def_ck}. 

If instead of the extrinsic curvature itself, the deformation is a cosine of the standard extrinsic curvature contraction $\K=6\H^2$ as defined in ref.~\cite{cuttellaction}, then ${\beta=\cos{\beta_k\K}\sim\cos{h^2}}$, which still cannot be transformed analytically.  However, it does match the function ${\beta(h)=1-4\pi_{\circ}^{-2}h^4}$ well, as we have plotted in \figref{fig:cosmo_cosine_def2}.  However, numerically finding the constraint for these two deformations, then considering the low $\R$ limit, we see that ${C\sim\R^2+C_\0}$.  Therefore, this deformation can be ruled out if ${C\sim\R+C_\0}$ is known to be the low curvature limit of the Hamiltonian constraint.

Returning to considering the function ${\beta(h)=1-4\pi_{\circ}^{-2}h^2}$ in \figref{fig:cosmo_cosine_def}, transforming from $h$ to $\K$ and from $r$ to $\R$ to $R$, we can see the correspondence between different limits of the deformation function,
\begin{equation}
    \beta \left( \K, 0 \right) = 1 - \beta_2 \K,
\quad \to \quad
    \beta \left( 0, R \right) = \frac{ 1 }{ 1 + \beta_2 R }.
\end{equation}
This is what we found in ref.~\cite{cuttellaction}, where the general form of this particular deformation is actually the product of these two limits.  However, for non-linear deformation functions, $\beta(\K,R)$ cannot be determined so easily from $\beta(\K,0)$ and $\beta(0,R)$. 
That being said, given $\beta(\R)$, the dependence on $\K$ can be found by solving the equations of motion.

%%%%%%%%%%%%%%%%%%%%%%%%%%%%%%%%%%%%%%%%%%%%%%%%%%%%%%%%%%

\section{Conclusions}
\label{sec:conclusions}

In this paper, we have found the general form that a deformed constraint can take for non-minimally coupled scalar-tensor variables.  The momenta and spatial derivatives for all fields must maintain the same relative structure in how they appear compared to the minimally-deformed constraint.  This means that the constraint is a function of the fields and the general kinetic term $\R$.  The freedom within this kinetic term comes down to the coupling functions.

While a lapse function transformation can apparently take the constraint algebra back to the undeformed form, this seems to be merely a cosmetic change as it does not in fact alter the Hamiltonian constraint itself.

We have shown how to obtain the cosmological equations of motion, and given a few simple examples of how they are modified.  For some deformation functions, a upper bound on energy density appears, which probably generates a cosmological bounce.  For other deformation functions, a sudden singularity in the expansion appears when the deformation diverges for high densities.  We have shown that deformations to the dynamics requires a background general potential against which the deformation can be contrasted.

Using the cosmological equations of motion, we made contact with the holonomy-generated deformation which is a cosine of the extrinsic curvature.  Through this, we have demonstrated how the relationship of momenta and extrinsic curvature becomes non-linear with a non-trivial deformation.  It seems that when the deformation produces an upper bound on extrinsic curvature, there does not seem to be an upper bound on intrinsic curvature or momenta.
Interestingly, the deformations which cause a big bounce seem to be required to vanish, but are not required to change sign.

One of the original motivations of this study was to provide insight into the problem of incorporating spatial derivatives, local degrees of freedom and matter fields into models of loop quantum cosmology which deform space-time covariance.  From our results, it would seem that the problem comes from considering the kinetic terms as separable, or as differently deformed.
The kinetic term, when constructed with canonical variables, cannot have its internal structure deformed beyond a sign.  The deformation can only be a function of the combined term, which means that matter field derivatives deform the space-time covariance in a similar way to curvature.
This may strike at the heart of the way the loop quantisation project, which attempts to first find a quantum theory of gravity, typically adds in matter as an afterthought.

That being said, there are important caveats to this work which must be kept in mind.  The fact that we used metric variables rather than the preferred connection or loop variables might limit the applicability of our results when comparing to the motivating theory.  Moreover, the deformation of the constraint algebra is only predicted for real values of $\BI$.
We also only considered combinations of derivatives or momenta that were a maximum of two orders, when higher order combinations and higher order derivatives are likely to appear in true quantum corrections.

%%%%%%%%%%%%%%%%%%%%%%%%%%%%%%%%%%%%%%%%%%%%%%%%%%%%%%%%%%

\section*{Acknowledgements}
\label{sec:ackowledgements}

We would like to thank Martin Bojowald for invaluable help during this study.
MS is supported in part by the Science and Technology Facility Council (STFC), UK under the research grant ST/P000258/1.
The work of RC is supported by an STFC studentship.

%%%%%%%%%%%%%%%%%%%%%%%%%%%%%%%%%%%%%%%%%%%%%%%%%%%%%%%%%%
%%%%%%%%%%%%%%%%%%%%%%%%%%%%%%%%%%%%%%%%%%%%%%%%%%%%%%%%%%

\appendix % Cue to tell LaTeX that the following "chapters" are Appendices

%%%%%%%%%%%%%%%%%%%%%%%%%%%%%%%%%%%%%%%%%%%%%%%%%%%%%%%%%%

\section{Decomposing the curvature}
\label{sec:curvature}

In our calculations, we need to decompose the three dimensional Riemann curvature frequently, so we collect the relevant identities in this appendix.  
We use the convenient definitions 
${\delta^{ab}_{cd}:=\delta^{(a}_{(c}\delta^{b)}_{d)}}$ and 
${Q_{abcd}:=q_{a(c}q_{d)b}}$.

The Riemann tensor is defined as the commutator of two covariant derivatives of a vector
\begin{equation}
    \nabla_c \nabla_d A^a - \nabla_d \nabla_c A^a 
    = R^a_{\;\;bcd} A^b,
\end{equation}
and can be given in terms of the Christoffel symbols,
\begin{equation}
    R^a_{\;\;bcd} = 
    \partial_c \Gamma^a_{db} - \partial_d \Gamma^a_{cb}
    + \Gamma^a_{ce} \Gamma^e_{db} - \Gamma^a_{ce} \Gamma^e_{cb},
\end{equation}
which are given by 
\begin{equation}
    \Gamma^a_{bc} = q^{ad} \partial_{(b} q_{c)d} - \half \partial^a q_{bc},
\end{equation}

The variation of the Riemann tensor is given by the Palatini equation,
\begin{equation}
	\delta R^a_{\;\,bcd} = \nabla_c \delta \Gamma^a_{db} - \nabla_d \delta \Gamma^a_{cb},
	    \label{eq:var_riemann_1}
\end{equation}
where the variation of the connection is
\begin{equation}
	\delta \Gamma^a_{bc} = 
	q^{ad} \nabla_{(b} \delta q_{c)d}
	- \half \nabla^a \delta q_{bc},
	    \label{eq:var_connection}
\end{equation}
from which we can calculate,
\begin{equation}
	\delta R^a_{\;\,bcd} = 
	\Theta^{a \;\;\;\;\; ef}_{\;\,bcd} \, \delta q_{ef}
	+ \Phi^{a \;\;\;\;\; efgh}_{\;\,bcd} \, \nabla_{e} \nabla_{f} \delta q_{gh}
	    \label{eq:var_riemann_2}
\end{equation}
where we've defined the useful tensors,
\begin{subequations}
\begin{align}
	\Theta^{a \;\;\;\;\; ef}_{\;\;bcd} & =
	\frac{-1}{2} \left( q^{a(e} R^{f)}_{\;\;\;\;bcd} + \delta^{(e}_b R^{f)a}_{\;\;\;\;\;\;\;cd} \right),
	    \label{eq:var_coeff_Theta} \\
	\Phi^{a \;\;\;\;\; efgh}_{\;\;bcd} & = \half \left( q^{a(e} \delta^{f)}_d \delta^{gh}_{bc} + q^{a(g} \delta^{h)}_d \delta^{ef}_{bc} - q^{a(e} \delta^{f)}_c \delta^{gh}_{bd} - q^{a(g} \delta^{h)}_c \delta^{ef}_{bd} \right),
	    \label{eq:var_coeff_Phi}
\end{align}
    \label{eq:var_coeff}%
\end{subequations}
\!\!but contracted versions of these are more useful,
\begin{subequations}
\begin{align}
	\Theta^{cd}_{ab}
	& := \delta^{ef}_{ab} \Theta^{g \;\;\;\;\; cd}_{\;\;egf} = \half \left( Q^{cdef} R_{e(ab)f} + \delta^{(c}_{(a} R^{d)}_{b)}	\right),
	\quad q_{cd} \Theta^{cd}_{ab} = q^{ab} \Theta^{cd}_{ab} = 0,
	    \label{eq:var_coeff_contract_Theta} \\
	\Phi_{ab}^{cdef}
	& := \delta^{gh}_{ab} \Phi^{i \;\;\;\;\; cdef}_{\;\;gih} = \half \left( q^{c(e} \delta^{f)d}_{ab} + q^{d(e} \delta^{f)c}_{ab} - q^{cd} \delta^{ef}_{ab} - q^{ef} \delta^{cd}_{ab} \right),
	    \label{eq:var_coeff_contract_Phi} \\
	\Phi^{abcd}
	& := q^{ef} \Phi^{abcd}_{ef} = Q^{abcd} - q^{ab} q^{cd}.
	    \label{eq:var_coeff_dblcontract_Phi}
\end{align}%
    \label{eq:var_coeff_contract}%
\end{subequations}
To decompose the Riemann tensor in terms of partial derivatives, use this formula for decomposing the second covariant derivative of the variation of the metric,
\begin{equation}
\begin{split}
    \nabla_d \nabla_c \delta q_{ab}
    & = \partial_d \partial_c \delta q_{ab} + \partial_g \delta q_{ef} \left( - \Gamma^g_{dc} \delta^{ef}_{ab} - 4 \delta^{(e}_{(a} \Gamma^{f)}_{b)(c} \delta^g_{d)} \right)
        \\
    &  + \delta q_{ef} \left( - 2 \partial_d \Gamma^{(e}_{c(a} \delta^{f)}_{b)} + 2 \Gamma^g_{dc} \Gamma^{(e}_{g(a} \delta^{f)}_{b)} + 2 \Gamma^g_{d(a} \delta^{(e}_{b)} \Gamma^{f)}_{cg} + 2 \Gamma^{(e}_{d(a} \Gamma^{f)}_{b)c} \right).
\end{split}
    \label{eq:var_metric}
\end{equation}

The two equations we need most are the derivative of the Ricci scalar with respect to the first and second spatial derivative of the metric, and we can find these from combining the above equations,
\begin{subequations}
\begin{align}
    \partdif{ R }{ \left( \partial_d \partial_c q_{ab} \right) }
    & =
    \partdif{ \left( \nabla_h \nabla_g q_{ef} \right) }{ \left( \partial_d \partial_c q_{ab} \right)}
    \partdif{ R }{ \left( \nabla_h \nabla_g q_{ef} \right) }
    = \delta_h^d \delta_g^c \delta_{ef}^{ab} \Phi^{efgh} = \Phi^{abcd}
\nonumber \\
    \therefore \partdif{ R }{ q_{ab,cd} } & = 
    \Phi^{abcd} = Q^{abcd} - q^{ab} q^{cd},
\\
    \partdif{ R }{ \left( \partial_c q_{ab} \right) }
    & =
    \partdif{ \left( \nabla_h \nabla_g q_{ef} \right) }{ \left( \partial_c q_{ab} \right)}
    \partdif{ R }{ \left( \nabla_h \nabla_g q_{ef} \right) }
    = \left( 
        - \Gamma^c_{gh} \delta^{ab}_{ef}
        - 4 \delta^{(a}_{(e} \Gamma^{b)}_{f)(g} \delta^c_{h)} 
    \right) \Phi^{efgh},
\nonumber \\
\begin{split}
    \therefore \partdif{ R }{ q_{ab,c} } & =
        \frac{3}{2} Q^{abde} \partial^c q_{de} 
        - Q^{edc(a} \partial^{b)} q_{de}
\\ &
        + q^{ab} Y^c
        - 2 q^{c(b} Y^{a)}
        - \half q^{ab} X^c
        + q^{c(b} X^{a)},
\end{split}
\end{align}
\end{subequations}
where 
$ X_a : = q^{bc} \partial_a q_{bc} $
and
$ Y_a : = q^{bc} \partial_{(c} q_{b)a} = \partial^b q_{ba} $.

%%%%%%%%%%%%%%%%%%%%%%%%%%%%%%%%%%%%%%%%%%%%%%%%%%%%%%%%%%

\section{The general diffeomorphism constraint}
\label{sec:diff}

We start from the assumption that the equal-time slices of our foliation are internally diffeomorphism covariant.  That is to say that spatial transformations and distortions are not deformed by the deformation of the constraint algebra.  As such, the Hamiltonian constraint is susceptible to deformation and the diffeomorphism constraint is not.  Therefore we need to consider what form the diffeomorphism constraint has.
In the hyperspace deformation algebra \eqref{eq:con-alg}, the diffeomorphism constraint forms a closed sub-algebra,
\begin{equation}
    \{ D_a [N^a], D_b [ M^b ] \} = D_a [ \mathcal{L}_M N^a].
\end{equation}
This equation shows that the diffeomorphism constraint is the generator of spatial diffeomorphisms (hence the name),
\begin{equation}
    \{ F, D_a [ N^a ] \} = \mathcal{L}_N F,
    \label{eq:diff_translation}
\end{equation}
for any phase space function $F$.  Using this relation, one can determine the unique form of the constraint for any field content

%%%%%%%%%%%%%%%%%%%%%%%%%%%%%%%%%%%%%%%%%5

Consider a scalar field $\left(\psi,\pi\right)$, and test \eqref{eq:diff_translation} using $F=\psi$,
\begin{subequations}
\begin{align}
\begin{split}
    \{ \psi (x), D_a [ N^a ] \}
    & =
    \int \mathrm{d}^3 y N^a (y) \funcdif{ D_a (y) }{ \pi (x) },
\\ &
    = N^a \partdif{ D_a }{ \pi } 
    - \partial_b \left( N^a \partdif{ D_a }{ \pi_{,b} } \right)
    + \partial_{bc} \left( N^a \partdif{ D_a }{ \pi_{,bc} } \right)
    + \ldots
\\ & 
    = N^a \left\{ 
        \partdif{ D_a }{ \pi } 
        - \partial_b \left( \partdif{ D_a }{ \pi_{,b} } \right) 
        + \partial_{bc} \left( \partdif{ D_a }{ \pi_{,bc} } \right)
    \right\}
\\ &
    + \partial_b N^a \left\{ 
        - \partdif{ D_a }{ \pi_{,b} }
        + 2 \partial_c \left( \partdif{ D_a }{ \pi_{,bc} } \right)
    \right\}
    + \partial_{bc} N^a \left( \partdif{ D_a }{ \pi_{,bc} } \right)
    + \ldots,
\end{split}
\\
    \mathcal{L}_N \psi & = N^a \partial_a \psi,
\end{align}%
\end{subequations}
comparing these two equations, one can easily see that 
\begin{equation}
    \partdif{ D_a }{ \pi } = \partial_a \psi,
\quad
    \partdif{ D_a }{ \pi_{,b} } = 0,
\quad
    \partdif{ D_a }{ \pi_{,bc} } = 0.
\end{equation}
If we check for any other $F(\psi,\pi)$, we find the same equations.
Therefore the diffeomorphism constraint for a scalar field is given by,
\begin{equation}
    D_a = \pi \partial_a \psi .
        \label{eq:diff_scalar}
\end{equation}

%%%%%%%%%%%%%%%%%%%%%%%%%%%%%%%%%%%%%%%%%%%%%%%%%%%%

If we consider a rank-2 spatial tensor ${(q_{ab},p^{cd})}$, and test \eqref{eq:diff_translation} again, we find the diffeomorphism constraint is given by
\begin{equation}
    D_a = p^{bc} \partial_a q_{bc} - 2 \partial_{(c} \left( q_{b)a} p^{bc} \right),
    \label{eq:diff_tensor}
\end{equation}
and for the specific example of a metric tensor, this reduces to 
\begin{equation}
    D_a = - 2 q_{ab} \nabla_c p^{bc}.
    \label{eq:diff_metric}
\end{equation}

%%%%%%%%%%%%%%%%%%%%%%%%%%%%%%%%%%%%%%

\begin{sloppypar}
For the general case of a tensor density%
\footnote{A tensor density does not transform as a tensor under changes of coordinates, but as a tensor multiplied by ${\left(\det{q_{ab}}\right)^{w/2}}$, where $w$ is the `weight' of the tensor density%
\cite{bojowald2010canonical}.}
with $n$ covariant indices, $m$ contravariant indices and weight $w$,
$ \left(
\Psi_{ a_1 \cdots a_n}^{ b_1 \cdots b_m},
\Pi^{ c_1 \cdots c_n}_{ d_1 \cdots d_m}
\right)$
where the canonical momentum has weight $1-w$,
the associated diffeomorphism constraint is given by,
\end{sloppypar}
\begin{equation}
    D_a = 
    \Pi^{ b_1 \cdots b_n}_{ c_1 \cdots c_m} \partial_a 
    \Psi_{ b_1 \cdots b_n}^{ c_1 \cdots c_m}
    - w \, \partial_a \left( \Pi^{ b_1 \cdots b_n}_{ c_1 \cdots c_m}
    \Psi_{ b_1 \cdots b_n}^{ c_1 \cdots c_m} \right)
    - n \, \partial_{(b_1} \left( 
        \Psi_{ b_2 \cdots b_n) a }^{ c_1 \cdots c_m } 
        \Pi^{ b_1 \cdots b_n}_{ c_1 \cdots c_m} 
    \right)
    + m \, \partial_{(c_1} \left( 
        \Pi_{ c_2 \cdots c_m) a }^{ b_1 \cdots b_n } 
        \Psi_{ b_1 \cdots b_n}^{ c_1 \cdots c_m}
    \right).
        \label{eq:diff_general}
\end{equation}

%%%%%%%%%%%%%%%%%%%%%%%%%%%%%%%%%%%%%%%%%%%%%%%%%%%%%%%%%%

\section{Extras}
\label{sec:allst_extras}

Use the following definitions for convenience,
\begin{equation}
    X_a = q^{bc} \partial_a q_{bc},
\quad
    Y_a = q^{bc} \partial_c q_{ba} = \partial^b q_{ab},
\quad
    Z_a = p_\T^{bc} \partial_a q_{bc},
\quad
    W_a = p_\T^{bc} \partial_c q_{ba}.
        \label{eq:allst_tensor_combinations}
\end{equation}
Evaluating each term in the $\partial_{cd}\theta^{ab}$ bracket of \eqref{eq:allst_dist-eqn-sol-metric},
\begin{subequations}
\begin{gather}
\begin{split}
    \partdif{ C }{ q_{ef,cd} } \partdif{^2 C }{ p^{ab} \partial p^{cd} }
    & = 2 \delta^{cd}_{ab} \partdif{ C }{ R } \partdif{ C }{ \bp }
    - 2 q_{ab} q^{cd} \partdif{ C }{ R } \left( \partdif{^2 C }{ p^2 } + \third \partdif{ C }{ \bp } \right)
    + 2 \left( q_{ab} p_\T^{cd} - 2 p^\T_{ab} q^{cd} \right) \partdif{ C }{ R } \partdif{^2 C }{ p \partial \bp }
\\ &
    + 4 p^\T_{ab} p_\T^{cd} \partdif{ C }{ R } \partdif{^2 C }{ \bp^2 }
\end{split}
\\
    - \partdif{ C }{ p^{ef } } \partdif{^2 C }{ q_{ef,cd} \partial p^{ab} }
    = 2 \left( 
        q^{cd} \partdif{ C }{ p } - p_\T^{cd} \partdif{ C }{ \bp } 
    \right) \left( 
        q_{ab} \partdif{^2 C }{ p \partial R }
    + 2 p^\T_{ab} \partdif{^2 C }{ \bp \partial R } 
    \right)
\\
    \partdif{ C }{ \psi_{,cd} }
    \partdif{^2 C }{ p^{ab} \partial \pi }
    = q^{cd} \partdif{ C }{ \Delta } \left(
        q_{ab} \partdif{^2 C }{ p \partial \pi }
        + 2 p^\T_{ab} \partdif{^2 C }{ \bp \partial \pi }
    \right)
\\
    - \partdif{ C }{ \pi } \partdif{^2 C }{ \psi_{,cd}} \partial p^{ab}
    = - \partdif{ C }{ \pi } q^{cd} \left(
        q_{ab} \partdif{^2 C }{ p \partial \Delta } + 2 p^\T_{ab} \partdif{^2 C }{ \bp \partial \Delta }
    \right)
\\
    - \beta \partdif{ D^c }{ p^{ab}_{,d} }
    = 2 \beta \delta_{ab}^{cd},
\end{gather}
    \label{eq:allst_extras_d2theta}%
\end{subequations}

Evaluating each term in the $\partial_c\theta^{ab}$ bracket of \eqref{eq:allst_dist-eqn-sol-metric},
\begin{subequations}
\begin{equation}
\begin{gathered}
    \partdif{ C }{ q_{ef,c} } \partdif{^2 C }{ p^{ab} \partial p^{ef} } = 
    \partdif{ C }{ \Delta }
    \left\{ q_{ab} \left[ 
        \partial^c \psi \left( \half \partdif{^2 C }{ p^2 } + \frac{2}{3} \partdif{ C }{ \bp } \right)
        - 2 p_\T^{cd} \partial_d \psi \partdif{^2 C }{ p \partial \bp }
    \right]
    - 2 \delta^c_{(a} \partial_{b)} \psi \partdif{ C }{ \bp }
\right. \\ \left.
    + p^\T_{ab} \left[
        \partial^c \psi \partdif{^2 C }{ p \partial \bp }
        - 4 p_\T^{cd} \partial_d \psi \partdif{^2 C }{ \bp^2 }
    \right]
    \right\}
   + \partdif{ C }{ R } \left\{
    \partdif{ C }{ \bp } \left[
        3 \partial^c q_{ab} 
        - 2 q^{cd} \partial_{(a} q_{b)d} 
        + 2 q_{ab} Y^c
        - q_{ab} X^c
        - 4 \delta^c_{(a} Y_{b)}
        + 2 \delta^c_{(a} X_{b)}
    \right]
\right. \\ \left.
    + q_{ab} X^c \left( \partdif{^2 C }{ p^2 } - \frac{2}{3} \partdif{ C }{ \bp } \right)
    + 2 p^\T_{ab} X^c \partdif{^2 C }{ p \partial \bp }
    + \left( 3 Z^c - 2 W^c - 4 p_\T^{cd} Y_d + 2 p_\T^{cd} X_d \right) \left(
        q_{ab} \partdif{^2 C }{ p \partial \bp }
        + 2 p^\T_{ab} \partdif{^2 C }{ \bp^2 } 
    \right)
    \right\},
\end{gathered}
\end{equation}
\begin{equation}
\begin{gathered}
    \partdif{^2 C }{ q_{ef,cd} } \partial_d \left( \partdif{^2 C }{ p^{ab} \partial p^{ef} } \right) =
    \partdif{ C }{ R } \left\{ 
        \left[ 
            q_{ab} \left( Y^c - X^c - 2 \partial^c \right)
            - 2 \partial^c q_{ab}
        \right] \left( \partdif{^2 C }{ p^2 } \! - \! \frac{2}{3} \partdif{ C }{ \bp } \right)
\right. \\ \left.
    + 2 \left( 
        \delta^c_{(a} \partial_{b)}
        - q_{ab} \partial^c
        + q^{cd} \partial_{(a} q_{b)d}
        + \delta^c_{(a} Y_{b)}
        - 2 \partial^c q_{ab}
    \right) \partdif{ C }{ \bp }
    + 2 \left[ 
        q_{ab} \left( 
            p_\T^{cd} \partial_d 
            + \partial_d p_\T^{cd}
            + W^c - Z^c
            + p_\T^{cd} Y_d
        \right)
\right. \right. \\ \left. \left.
        + p_\T^{cd} \partial_d q_{ab}
        + p^\T_{ab} \left(
            Y^c - X^c - 2 \partial^c
        \right)
        - 2 Q_{abde} \partial^c p_\T^{de}
        - 4 \partial^c q_{d(a} p_{\;b)}^{\T\,d}
    \right] \partdif{^2 C }{ p \partial \bp }
\right. \\ \left.
    + 4 \left[ 
        Q_{abef} p_\T^{cd} \partial_d p_\T^{ef}
        + 2 p_\T^{cd} \partial_d q_{e(a} p_{\;b)}^{\T\,e}
        + p^\T_{ab} \left( 
            \partial_d p_\T^{cd}
            + W^c - Z^c + p_\T^{cd} Y_d + p_\T^{cd} \partial_d
        \right)
    \right] \partdif{^2 C }{ \bp^2 }
    \right\}
\end{gathered}
\end{equation}
\begin{equation}
\begin{gathered}
    \partdif{ C }{ p^{ef} } \partdif{^2 C }{ q_{ef,c} \partial p^{ab} } =
    2 p_\T^{cd} \partdif{ C }{ \bp } \left\{
        q_{ab} \partial_d \partdif{^2 C }{ p \partial R }
        + 2 p^\T_{ab} \partial_d \partdif{^2 C }{ \bp \partial R }
    \right\}
\\
    + \left\{ \partdif{ C }{ p } \left[
        X^c \partdif{}{ R } 
        + \half \partial^c \psi \partdif{}{ \Delta }
    \right]
    + \partdif{ C }{ \bp } \left[ 
        \left( 3 Z^c - 2 W^c - 4 p_\T^{cd} Y_d \right) \partdif{}{ R }
        - 2 p_\T^{cd} \partial_d \psi \partdif{}{ \Delta }
    \right]
    \right\} \left( q_{ab} \partdif{ C }{ p } + 2 p^\T_{ab} \partdif{ C }{ \bp } \right)
    ,
\end{gathered}
\end{equation}
\begin{equation}
\begin{gathered}
    \partdif{ C }{ p^{ef} } \partial_d \left( \partdif{^2 C }{ q_{ef,cd} \partial p^{ab} } \right) =
    2 \partdif{ C }{ \bp } \left\{ 
        \left[
            q_{ab} \left( Z^c \! - \! W^c \! - \! p_\T^{cd} Y_d \! + \! p_\T^{cd} \partial_d \right)
            + p_\T^{cd} \partial_d q_{ab}
        \right] \partdif{^2 C }{ p \partial R }
\right. \\ \left.
        + 2 \left[
            p^\T_{ab} \left( 
                Z^c \! - \! W^c \! - \! p_\T^{cd} Y_d \! + \! p_\T^{cd} \partial_d
            \right)
            + Q_{abef} p_\T^{cd} \partial_d p_\T^{ef}
            + 2 p_\T^{cd} \partial_d q_{e(a} p_{\;b)}^{\T\,e}
        \right] \partdif{^2 C }{ \bp \partial R }
    \right\}
\\
    +\partdif{ C }{ p } \left\{ 
        \left[
            q_{ab} \left( X^c + Y^c - 2 \partial^c \right)
            - 2 \partial^c q_{ab}
        \right] \partdif{^2 C }{ p \partial R }
        + 2 \left[
            p^\T_{ab} \left( X^c + Y^c - 2 \partial^c \right)
            - 2 Q_{abef} \partial^c p_\T^{ef}
            - 4 \partial^c q_{d(a} p^{\T\,d}_{\;b)}
        \right] \partdif{^2 C }{ \bp \partial R }
    \right\}
\end{gathered}
\end{equation}
\begin{equation}
    \partdif{ C }{ \psi_{,c} } \partdif{^2 C }{ \pi \partial p^{ab} } =
    \left\{ 
        2 \partial^c \psi \partdif{ C }{ \gamma }
        + \left( \half X^c - Y^c \right) \partdif{ C }{ \Delta }
    \right\} \left( 
        q_{ab} \partdif{ C }{ \pi \partial p }
        + 2 p^\T_{ab} \partdif{ C }{ \pi \partial \bp }
    \right),
\end{equation}
\begin{equation}
\begin{split}
    \partdif{ C }{ \psi_{,cd} } \partial_d \left( \partdif{^2 C }{ \pi \partial p^{ab} } \right)
    & =
    \partdif{ C }{ \Delta } \left\{ \left( 
            \partial^c q_{ab} + q_{ab} \partial^c 
        \right) \partdif{^2 C }{ \pi \partial p }
        + 2 \left(
            Q_{abef} \partial^c p_\T^{ef} 
            + 2 \partial^c q_{d(a} p_{\;a)}^{\T\,d}
            + p^\T_{ab} \partial^c
        \right) \partdif{^2 C }{ \pi \partial \bp }
    \right\},
\end{split}
\end{equation}
\begin{equation}
\begin{split}
    \partdif{ C }{ \pi } \partial_d \left( \partdif{^2 C }{ \psi_{,cd} \partial p^{ab} } \right)
    & =
    \partdif{ C }{ \pi } \left\{
        \left( 
            q_{ab} \partial^c + \partial^c q_{ab} - q_{ab} Y^c
        \right) \partdif{^2 C }{ p \partial \Delta }
\right. \\ & \left.
        + 2 \left(
            Q_{abef} \partial^c p_\T^{ef}
            + 2 \partial^c q_{d(a} p_{\;b)}^{\T\,d}
            + p^\T_{ab} \partial^c
            - p^\T_{ab} Y^c
        \right) \partdif{^2 C }{ \bp \partial \Delta }
    \right\},
\end{split}
\end{equation}
\begin{equation}
\begin{split}
    \partdif{ ( \beta D^c ) }{ p^{ab} }
    & = 
    \beta \left( \partial^c q_{ab} - 2 q^{cd} \partial_{(a} q_{b)d} \right)
    + \left( 
        q_{ab} \partdif{ \beta }{ p }
        + 2 p^\T_{ab} \partdif{ \beta }{ \bp }
    \right)
    \left(
        \pi \partial^c \psi
        - 2 \partial_d p_\T^{cd}
        - \frac{2}{3} \partial^c p
        - 2 W^c + Z^c
        + \third p X^c
    \right),
\end{split}
\end{equation}
\begin{equation}
    \partial_d \left( \beta \partdif{ D^c }{ p^{ab}_{,d} } \right)
    = - 2 \delta^c_{(a} \partial_{b)} \beta,
\end{equation}
    \label{eq:allst_extras_dtheta}%
\end{subequations}

Evaluating each term in the $\partial_{cd}\eta^{ab}$ bracket of \eqref{eq:allst_dist-eqn-sol-scalar},
\begin{subequations}
\begin{gather}
    \partdif{ C }{ q_{cd,ab} } \partdif{^2 C }{ \pi \partial p^{cd} } =
    \partdif{ C }{ R } \left(
        - 2 q^{ab} \partdif{^2 C }{ \pi \partial p }
        + 2 p_\T^{ab} \partdif{^2 C }{ \pi \partial \bp }
    \right),
\\
    \partdif{ C }{ p^{cd} } \partdif{^2 C }{ q_{cd,ab} \partial \pi } =
    \partdif{^2 C }{ R \partial \pi } \left(
        - 2 q^{ab} \partdif{ C }{ p }
        + 2 p_\T^{ab} \partdif{ C }{ \bp }
    \right),
\\
    \partdif{ C }{ \psi_{,ab} } \partdif{^2 C }{ \pi^2 }
    - \partdif{ C }{ \pi } \partdif{^2 C }{ \psi_{,ab} \partial \pi }
    =
    q^{ab} \left( 
        \partdif{ C }{ \Delta } \partdif{^2 C }{ \pi^2 }
        - \partdif{ C }{ \pi } \partdif{^2 C }{ \Delta \partial \pi }
    \right),
\end{gather}
    \label{eq:allst_extras_d2eta}%
\end{subequations}

Evaluating each term in the $\partial_c\eta^{ab}$ bracket of \eqref{eq:allst_dist-eqn-sol-scalar},
\begin{subequations}
\begin{equation}
\begin{split}
    \partdif{ C }{ q_{cd,a} } \partdif{^2 C }{ \pi \partial p^{cd} } & =
    \partdif{ C }{ R } \left\{ 
        X^a \partdif{^2 C }{ \pi \partial p }
        + \left( 
            3 Z^a - 2 W^a - 4 p_\T^{ab} Y_b + 2 p_\T^{ab} \partial_b
        \right) \partdif{^2 C }{ \pi \partial \bp }
    \right\}
\\ &
    + \partdif{ C }{ \Delta } \left\{
        \half \partial^a \psi \partdif{^2 C }{ \pi \partial p }
        - 2 p_\T^{ab} \partial_b \psi \partdif{^2 C }{ \pi \partial \bp }
    \right\},
\end{split}
\end{equation}
\begin{equation}
\begin{split}
    \partdif{ C }{ q_{cd,ab} } \partial_b \left( \partdif{^2 C }{ \pi \partial p^{cd} } \right) & =
    \partdif{ C }{ R } \left\{ 
        \left( Y^a - X^a - 2 \partial^a 
        \right) \partdif{^2 C }{ \pi \partial p }
        + 2 \left( 
            \partial_b p_\T^{ab} + W^a - Z^a + p_\T^{ab} Y_b + p_\T^{ab} \partial_b
        \right) \partdif{^2 C }{ \pi \partial \bp }
    \right\},
\end{split}
\end{equation}
\begin{equation}
\begin{split}
    \partdif{ C }{ p^{cd } } \partdif{^2 C }{ q_{cd,a} \partial \pi } & =
    \partdif{ C }{ p } \left\{ 
        X^a \partdif{^2 C }{ R \partial \pi }
        + \half \partial^a \psi \partdif{^2 C }{ \Delta \partial \pi }
    \right\}
\\ &
    + \partdif{ C }{ \bp } \left\{
        \left( 
            3 Z^a - 2 W^a - 4 p_\T^{ab} Y_b + 2 p_\T^{ab} X_b
        \right) \partdif{^2 C }{ R \partial \pi }
        - 2 p_\T^{ab} \partial_b \psi \partdif{^2 C }{ \Delta \partial \pi }
    \right\},
\end{split}
\end{equation}
\begin{equation}
\begin{split}
    \partdif{ C }{ p^{cd} } \partial_b \left( \partdif{^2 C }{ q_{cd,ab} \partial \pi } \right) & =
    \left\{ 
        \partdif{ C }{ p } \left( X^a + Y^a - 2 \partial^a \right)
        + 2 \partdif{ C }{ \bp } \left( Z^a - W^a - p_\T^{ab} Y_b + p_\T^{ab} \partial_b \right)
    \right\} \partdif{^2 C }{ R \partial \pi },
\end{split}
\end{equation}
\begin{equation}
\begin{split}
    \partdif{ C }{ \psi_{,a} } \partdif{^2 C }{ \pi^2 }
    + \partdif{ C }{ \pi } \partdif{^2 C }{ \psi_{,a} \partial \pi }
    & = 
    2 \partial^a \psi \left(
        \partdif{ C }{ \gamma } \partdif{^2 C }{ \pi^2 }
        + \partdif{ C }{ \pi } \partdif{^2 C }{ \pi \partial \gamma }
    \right)
    + \left( \half X^a + Y^a \right) \left(
        \partdif{ C }{ \Delta } \partdif{^2 C }{ \pi^2 }
        + \partdif{ C }{ \pi } \partdif{^2 C }{ \pi \partial \Delta }
    \right)
\end{split}
\end{equation}
\begin{equation}
    \partdif{ C }{ \psi_{,ab} } \partial_b \left( \partdif{^2 C }{ \pi^2 } \right)
    - \partdif{ C }{ \pi } \partial_b \left( \partdif{^2 C }{ \psi_{,ab} \partial \pi } \right)
    = \partdif{ C }{ \Delta } \partial^a \left( \partdif{^2 C }{ \pi^2 } \right)
    + \partdif{ C }{ \pi } \left( Y^a - \partial^a \right) \partdif{^2 C }{ \pi \partial \Delta },
\end{equation}
\begin{equation}
    \partdif{ ( \beta D^a ) }{ \pi } =
    \partdif{ \beta }{ \pi } \left( 
        - 2 \partial_b p_\T^{ab} 
        - \frac{2}{3} \partial^a p
        + 2 W^a - Z^a - \third p X^a
    \right)
    + \partial^a \psi \left( \beta + \pi \partdif{ \beta }{ \pi } \right),
\end{equation}
    \label{eq:allst_extras_deta}%
\end{subequations}

%%%%%%%%%%%%%%%%%%%%%%%%%%%%%%%%%%%%%%%%%%%%%%%%%%%%%%%%%%

\bibliographystyle{./utphys}  % Use the "unsrtnat" BibTeX style for formatting the Bibliography
\bibliography{bibliography}  % The references (bibliography) information are stored in the file named "Bibliography.bib"

%%%%%%%%%%%%%%%%%%%%%%%%%%%%%%%%%%%%%%%%%%%%%%%%%%%%%%%%%%

\end{document}